\documentclass[manuscript]{aastex}
\usepackage{soul,color}
\usepackage{subfloat}

\usepackage{graphicx}
\usepackage{romannum}

\slugcomment{}

\shorttitle{X-ray Polarization of Coronal Emission in AGNs }
\shortauthors{Beheshtipour et al.}

\begin{document}

\title{The X-ray Polarization of the Accretion Disk Coronae of Active Galactic Nuclei}

\author{Banafsheh Beheshtipour\altaffilmark{1}, Henric Krawczynski\altaffilmark{1}, and Julien Malzac\altaffilmark{2}}

\affil{$^1$Physics Department and McDonnell Center for the Space Sciences, Washington University in St. Louis, One Brookings Drive, CB 1105, St. Louis, MO 63130, USA}
\affil{$^2$IRAP, Universit\'e de toulouse, CNRS, UPS, CNES , Toulouse, France}

\altaffiltext{1}{Email: b.beheshtipour@wustl.edu}

\begin{abstract}
Hard X-rays observed in Active Galactic Nuclei (AGNs) are thought to originate from the Comptonization of the optical/UV accretion disk photons in a hot corona. Polarization studies of these photons can help to constrain the corona geometry and the plasma properties. We have developed a ray-tracing code that simulates the Comptonization of accretion disk photons in coronae of arbitrary shape, and use it here to study the polarization of the X-ray emission from wedge and spherical coronae. We study the predicted polarization signatures for the fully relativistic and various approximate treatments of the elemental Compton scattering processes. We furthermore use the code to evaluate the impact of non-thermal electrons and cyclo-synchrotron photons on the polarization properties. Finally, we model the NuSTAR observations of the Seyfert I galaxy Mrk 335 and predict the associated polarization signal.  Our studies show that X-ray polarimetry missions such as NASA's Imaging X-ray Polarimetry Explorer (IXPE) and the X-ray Imaging Polarimetry Explorer (XIPE) proposed to ESA will provide valuable new information about the physical properties of the plasma close to the event horizon of AGN black holes.
\end{abstract}


\section{Introduction}
Active Galactic Nuclei (AGNs) are powerful sources of X-rays. Their spectrum is dominated by a power law continuum presumably emitted by hot and possibly partially non-thermal plasma of particles, known as a corona. Repeated inverse Compton processes in the corona energize  optical/UV photons originating from an accretion disk emitting in the IR/optical/UV. Even though the first accretion disk and corona models were developed in the seventies \citep[e.g][]{sha73,nov73} and refined over the last 40 years \citep[e.g.][]{haa91, dov97, now02}, the geometry of the corona, i.e. its location and spatial extent, is still a matter of intense debate \citep{gil14}.

Recent X-ray reverberation observations \citep[e.g.][]{fab09, wil13} and future X-ray polarization observations 
offer a new way of constraining the corona geometry  that is complementary to the more traditional constraints from X-ray spectroscopy. Polarimetric observations with the Imaging X-ray Polarimetry Explorer (IXPE) Small Explorer (SMEX) mission \citep{wei14}, and possibly with the proposed X-ray Imaging Polarimetry Explorer (XIPE) ESA mission \citep{sof13} promise to provide geometrical information about the corona and the inner structure of accretion disks. 
The polarization of X-ray emission from the accretion disk of a stellar mass BH is predicted to be linear polarized with the polarization fraction being a function of inclination of the disk \citep[and references there in]{li09}. In the case of AGNs, X-rays are emitting from a hot corona in the vicinity of the accretion disk. The polarization of this emission depends on the scattering off the accretion disk and on the scattering  processes in the corona. This dependency of the polarization on the scattering makes polarization studies a promising way to distinguish between different corona geometries. \citet{sch10} showed that in stellar mass BH, the corona geometry has a major impact on the predicted energy spectra of the polarization fraction and the polarization angle. \citet{dov12} studied the polarization of unpolarized corona X-rays scattering off the accretion disk of AGNs. In this paper, we study for the first time the impact of the Klein-Nishina (K-N, \citet{kle29}) cross section on the polarization of the coronal emission.  Furthermore, we study how non-thermal electrons in the coronal plasma and polarized synchrotron and cyclotron seed photons affect the observable polarization properties. Our studies are based on a general relativistic ray-tracing code that simulates the individual scattering processes accounting for the energy dependent K-N cross section in the framework of a general relativistic ray tracing code. The code assumes that the 3-D corona plasma orbits the black hole with the angular velocity of a ZAMO (Zero Angular Momentum Observer). The seed photons are polarized with an initial polarization given by the classical results of Chandrasekhar (1960). The code tracks photons forward in time, making it possible to study repeated scatterings in the corona and off the accretion disk. 

Although simple models assume a single-temperature corona, the coronal plasma may have a distribution of temperatures and/or an admixture of non-thermal plasma (e.g. from magnetic reconnection in the corona). The Atacama Large Millimeter/Submillimeter Array (ALMA) may be able to reveal the presence of non-thermal plasma in AGN coronae \citep{ino14}. The energy spectra of black holes in X-ray binaries show clear evidence for non-thermal particles  \citep[e.g.][]{cop90, pou98, gie99, rom14, mal16}. In Cyg X-1, the non-thermal component was detected to be strongly polarized \citep{jou12, lau11}, and this was taken as evidence that it is formed as synchrotron emission in the jet rather than as inverse Compton emission in the corona.  Further below, we will use our code to evaluate the possibility that non-thermal electrons in the corona produce high polarization fractions at high energies.

The structure and strength of the magnetic fields in AGN accretion disks is still a matter of debate \citep[e.g.][]{bel05, bla77}. Cyclotron or synchrotron (cyclo-synchrotron) photons are naturally expected from the energetic electrons radiating in the ambient coronal magnetic field \citep[see e.g.][]{mal09, vel11}. In this paper, we will evaluate if X-ray polarization could contribute to clarifying the situation if a fraction of the seed photons are highly polarized cyclotron or synchrotron photons. 

The rest of the paper is structured as follows. We describe the ray-tracing code and the corona geometries in Sect. 2. We report on the results of the studies of the impact of the K-N cross section, non-thermal electrons, and  cyclotron and synchrotron seed photons in Sections 3 to 5. In Section 6 we model the NuSTAR observations of Mrk 335, and use the modeling to predict the polarization fraction and angle energy spectra. We summarize our results in Sec. 7. Throughout this paper, all distances are given in units of the gravitational radius $r_{\rm g} =GM/c^2$, and we set $G=c=\hbar=1$. The mass of AGN is $10^8 M_{\sun}$ unless otherwise specified. The inclination is $i = 0^{\circ}$ for an observer viewing the disk face-on and $i = 90^{\circ}$ for an observer viewing the disk edge-on.

\section {Ray tracing}\label{sec2}
\subsection{Thermal disk simulation}
We expanded upon the general relativistic ray-tracing code described in \citet{kra12,beh16,hor16}, called the thermal code in the following. The code simulates an accretion disk extending from the innermost stable circular orbit (ISCO) $r_{\rm ISCO}$ to $r_{\rm max}=100\,r_g$. The disk emits thermally with a radial brightness distribution given by mass, energy, and angular momentum conservation \citep{pag74}. Our simulations ignore the emission and scattering by gas inside the ISCO. The code uses Boyer Lindquist coordinates and tracks photons forward in time using the fourth order Runge-Kutta method to integrate the geodesic equation and to parallel transport the polarization vector. All photons are tracked until they come to within 0.02\% $r_g$ to the event horizon at which point we assume they will cross the event horizon and will not contribute to the observed signal. The code assigns an initial polarization to each photon using Chandrasekhar's parameterization of the limb brightening and polarization of the emission from an indefinitely deep atmosphere of free electrons \citep[][Tables 24 and 25]{cha60}. The polarization fraction is Lorentz invariant and only changes when the photon scatters.

When a photon hits the disk, its wave and polarization four vectors are transformed from the global Boyer Lindquist coordinate system into the plasma rest frame, the polarization fraction and the polarization vector are used to calculate the Stokes parameters, a random new direction is drawn, Chandrasekhar's results for Thomson scattering off an indefinitely deep electron atmosphere are used to modify the Stokes parameters for the given incident and scattered direction, the Stokes {\it I} parameter is used to modify the statistical weight of the photon, the {\it I}, {\it Q} and {\it U} parameters are used to calculate the polarization fraction and vector, and the wave and polarization four vectors are back-transformed from the plasma rest frame into the global Boyer Lindquist coordinate system.

\begin{figure}
\epsscale{0.6}
\plotone{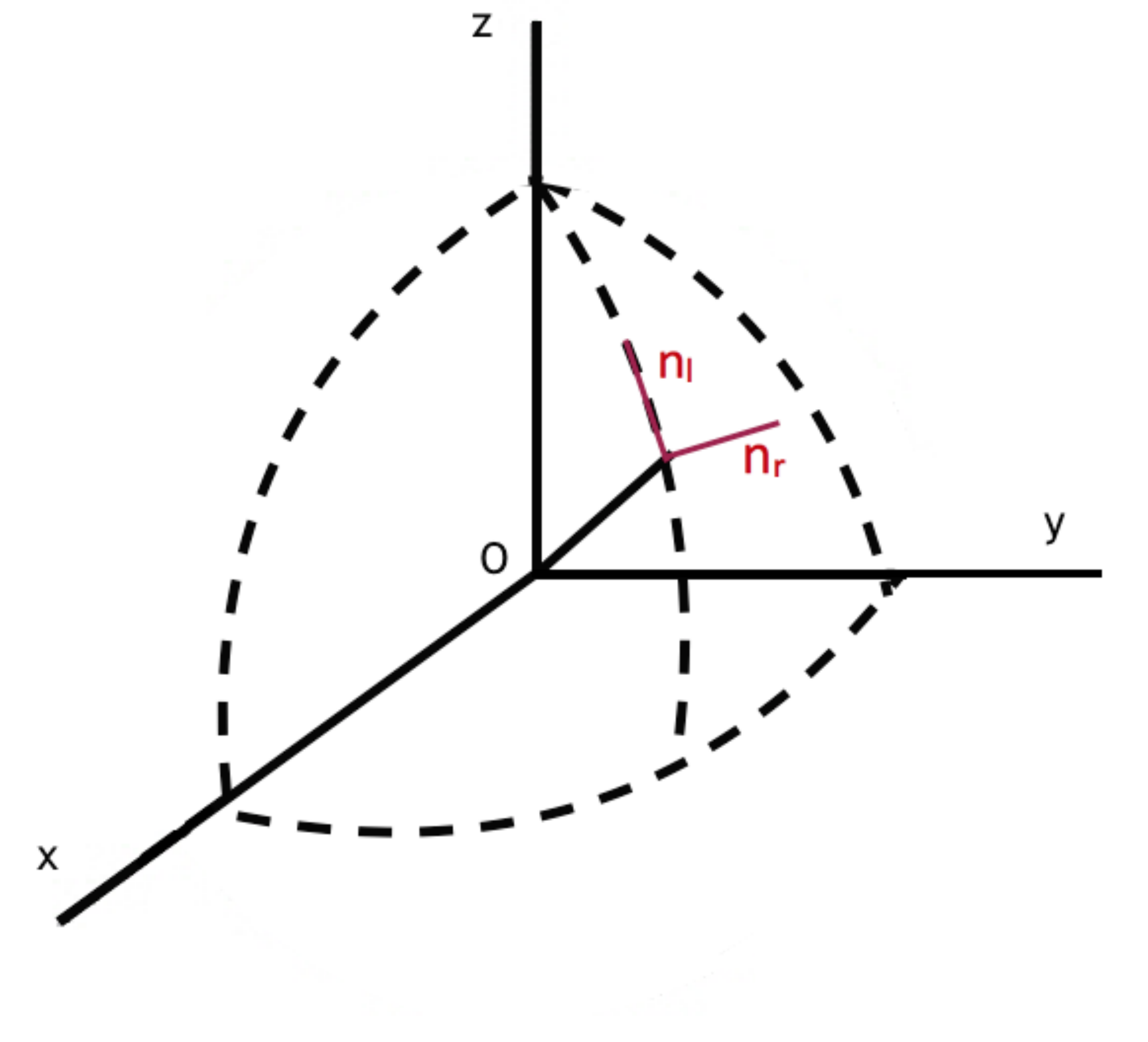} 
\caption{The meridian plane coordinate system used to derive polarization and Stokes parameters.}
\label{cor2}
\end{figure}

\citet[]{cha60} derived the following equation for the 
intensities $I_l(\tau,\mu)$ and $I_r(\tau,\mu)$ in the directions parallel
and perpendicular to the meridian plane (see Fig.~\ref{cor2}) 
and the Stokes parameter $U$ \citep[][Chapter 10, Equ. (163)]{cha60}):
\begin{equation}
I(0,\mu,\phi)=\left(
\begin{array}{cc}
I_l \\
I_r \\
U\\
\end{array} \right)=\frac{1}{4\mu}\bold{Q}\bold{S}(\mu,\phi;\mu_0,\phi_0)\left(
\begin{array}{cc}
F_l \\
F_r \\
F_U\\
\end{array} \right)
\end{equation}
where $\mu_0$ and $\mu$ are the direction cosines of the incident and scattered emission, $F_{\rm l}$ and $F_{\rm r}$ are the incident fluxes in the {\it{l}} and {\it{r}} directions, and $\bold{Q}$ and $\bold{S}$ are matrices tabulated in \citet{cha60}.
For each scattering, we multiply the statistical weight with 
$2\pi\,\eta\, \mu/\mu_0\,I/F$ (with $I=I_{\rm l}+I_{\rm r}$).
The factor $2\pi$ normalizes the expression to $\eta$ when averaging over all scattering directions with $\eta\le 1$ being the scattering efficiency ($\eta=1$ if not mentioned otherwise), and the factor $\mu/\mu_0$ normalizes $I$ and $F$ to the flux per accretion disk area. Note that the polarization fraction is given by $\Pi=|(I_{\rm r}-I_{\rm l})|/(I_{\rm r}+I_{\rm l})$. 

We track photons until they reach a Boyer Lindquist radial coordinate of $r\,=\,10,000\,r_{\rm g}$.
The photon wave and polarization vectors are subsequently transformed into the reference frame of an observer at fixed coordinates. We study inclination dependent effects by collecting photons arriving within 
$\pm 4^{\circ}$ of the $\theta$-angle of the observer.
We calculate the polarization fraction $\Pi$ and angle $\chi$ by summing the 
Stokes parameters of the individual photons and then using the standard equations:
\begin{equation}
\Pi\,=\,\sqrt{(U^{\,2}+Q^{\,2})/I^2}
\label{s1}
\end{equation}
\begin{equation}
\chi=1/2\tan^{-1}(U/Q).
\label{s2}
\end{equation}
with the summed Stokes parameters $I$, $Q$ and $U$.
We find that it is important to calculate statistical errors on the polarization results.
We do so by calculating the Stokes parameters for 11 independent subsets of the simulated data sets, 
and deriving 11 independent estimates of $I$, $Q$ and $U$.
The mean values and standard deviations of the three Stokes parameters are subsequently used
to calculate the mean values and standard deviations of the polarization fraction and angle
based on Equations (\ref{s1}) and (\ref{s2}) and standard error propagation.
\subsection{Comptonization in the Corona}
\begin{figure}
  \centering
  \begin{tabular}{@{}c@{}}
    \includegraphics[width=.7\linewidth,height=100pt]{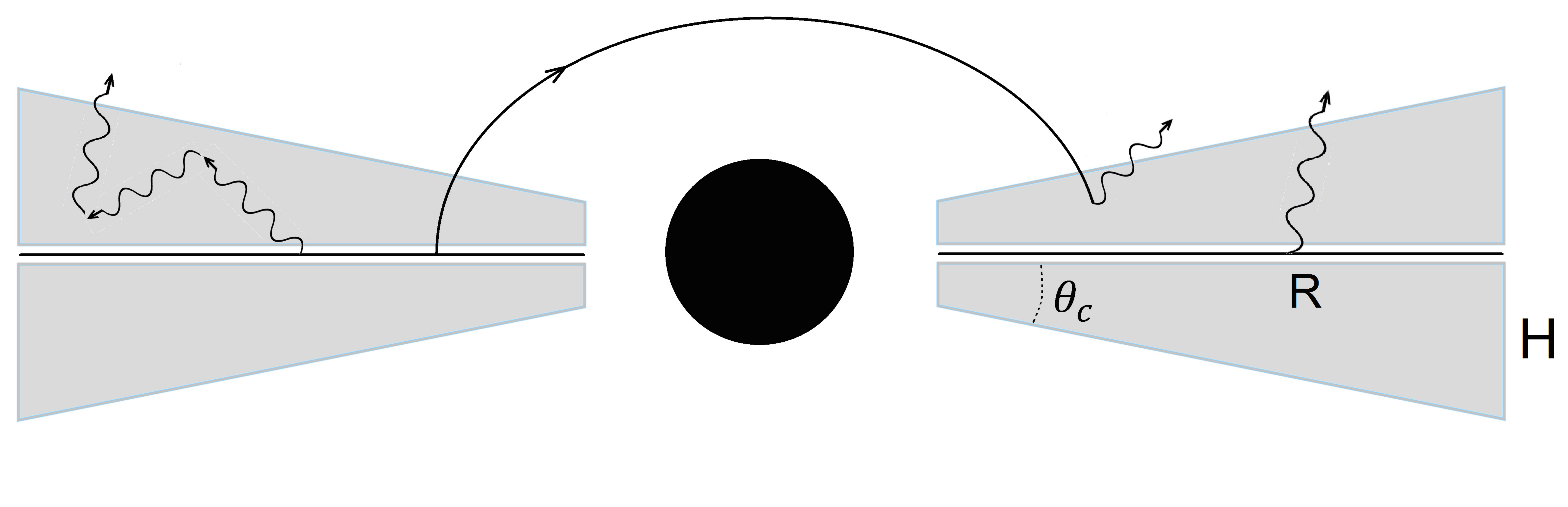} \\[\abovecaptionskip]
    \small (a) Wedge corona
  \end{tabular}

  \vspace{\floatsep}

  \begin{tabular}{@{}c@{}}
    \includegraphics[width=.7\linewidth,height=120pt]{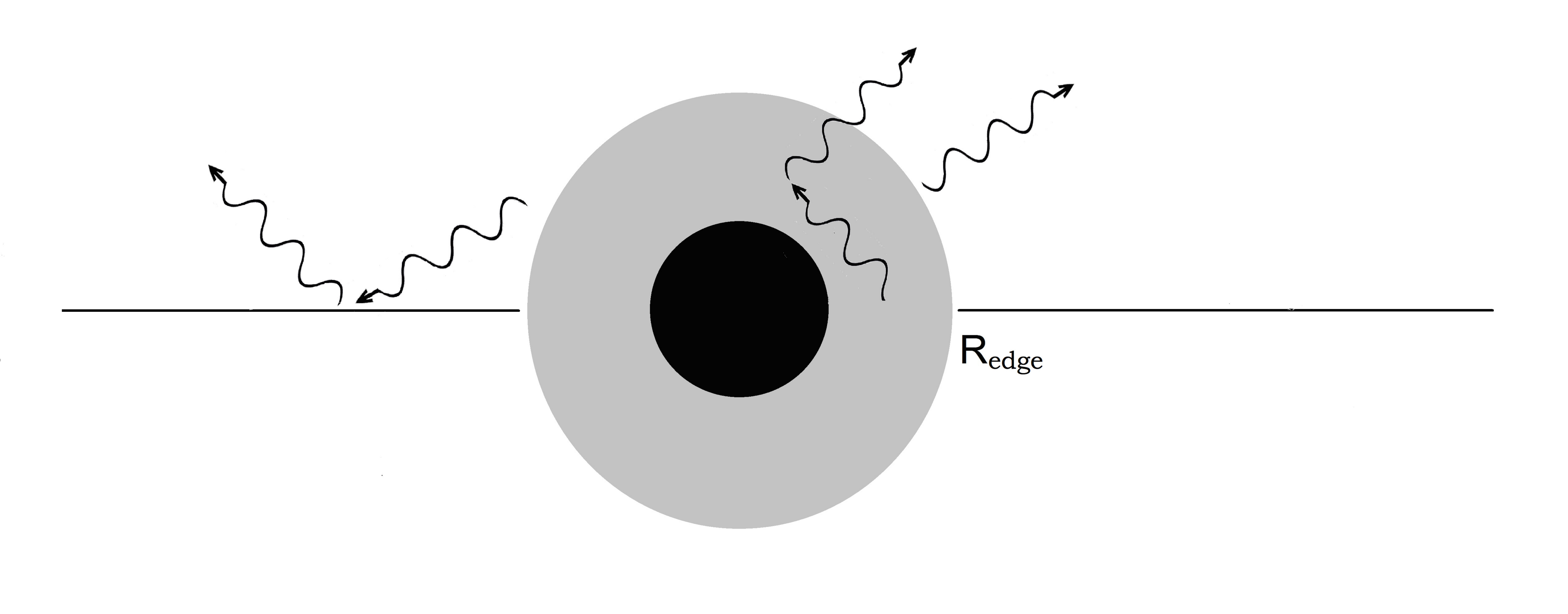} \\[\abovecaptionskip]
    \small (b) Spherical corona 
  \end{tabular}

  \caption{Sketch of the wedge and spherical corona geometries. (a) The wedge corona extends above and below the accretion disk with an opening angle of $\theta_c=\tan^{-1}H/R$. (b) The spherical corona extends from the black hole horizon to $R_{edge}$. The disk is truncated at $R_{edge}$.}\label{fig:tot}
\end{figure}

We simulate isothermal wedge and spherical coronae. 
The wedge coronae lie above and below the accretion disk and are chosen to have a constant opening angle of $8^{\circ}$ 
(see Figure~\ref{fig:tot} (a)). A photon traversing a distance $dl$ in the corona rest frame scatters with the probability 
$p \,=\,e^{-d\tau}$ with: 
\begin{equation}
d\tau(r,z)=\kappa \rho(r,z) dl
\end{equation}
where $\kappa\,=\,0.4\, cm^2/g$ is the opacity to electron scattering and $\rho(r,z)$ 
is the density of the coronal plasma at radius $r$ and height $z$.
For the wedge corona, we assume the coronal gas density dependence on radius and height from \citet[][Equation (6)]{sch10}:
\begin{equation}
\rho(r,z)=\rho_0(r) \exp(-z/H(r)), \quad
\rho_0(r)=\frac{\tau_0}{\kappa H(r)}
\end{equation}
with the adjustable parameter $\tau_0$. Since the total optical depth is a function of the worldline, 
we characterize the density with the scattering coefficient (the optical depth per proper length)  
$\sigma=d\tau/dl$, and we quote $\sigma$  for the mean height and the mean radius of the disk.

The spherical corona extends from $r_{\rm ISCO}$ to $R_{\rm edge}$. 
In this geometry the accretion disk is truncated at the outer edge of the corona at 
$R_{\rm edge}=15 r_{\rm g}$ (Fig.~\ref{fig:tot} (b)). The optical depth is chosen to be a 
linear function of radius, $\tau(r)=(\tau_0/R_c)r$ and $R_c=R_{edge}-r_{ISCO}$ is the radius of the corona and $\sigma$ is $\tau_0/R_c$. In the wedge corona, all seed photons are thermally emitted accretion disk photons. The disk of the spherical corona model is truncated at the edge of the corona, and we assume the same seed photon luminosity and energy spectrum as function of the radial coordinate $r$ as for the wedge corona, and launch the photons with a random polar angle $\theta$ with a flat $\cos{(\theta)}$ distribution.
 
The calculation assumes that the coronal plasma rotates with the ZAMO at an angular velocity of 
$v_{\phi}=-g_{\phi t}/g_{\phi\phi}$ with $g_{\phi\phi}$ and $g_{\phi t}$ being components of the Kerr metric. 
A random number is drawn to decide if the photon scatters. If it does, its wave vector and polarization vector 
are transformed first to the rest frame of the coronal plasma, and then to the electron rest frame 
where the new scattering energy and  polarization vector are calculated.  This process will be described in more detail in the next Section.  After the scattering, the photon wave vector and polarization vector are transformed back, first into the coronal rest frame (CR)  and then into the global Boyer-Lindquist frame, where the tracking continues.
\section{Thomson and K-N Scatterings}\label{kn}
We simulate photon-electrons scatterings using the Thomson approximation and the full K-N cross section. 
In both cases, we transform the wavevector and polarization vector of the photon 
first from the global BL coordinates into the corona frame coordinates, 
and subsequently into the rest frame of one of the scattering 
electrons (assumed to be isotropic in the corona frame). We randomly draw the direction of the scattered photon in the electron rest frame. In the Thomson approximation the scattering does not change the photon energy.
More accurately, the photon looses energy according to the Compton's equation:
\begin{equation}\label{a}
\epsilon_1=\frac{\epsilon_0}{1+x(1-{\rm cos}\theta)},
\end{equation}
where $\epsilon_1$ and $\epsilon_0$ are the energy of the photon after and before scattering, respectively, 
$x$ is the photon energy in units of the electron rest mass, and $\theta$ is the scattering angle.

\begin{figure}
\epsscale{0.6}
\plotone{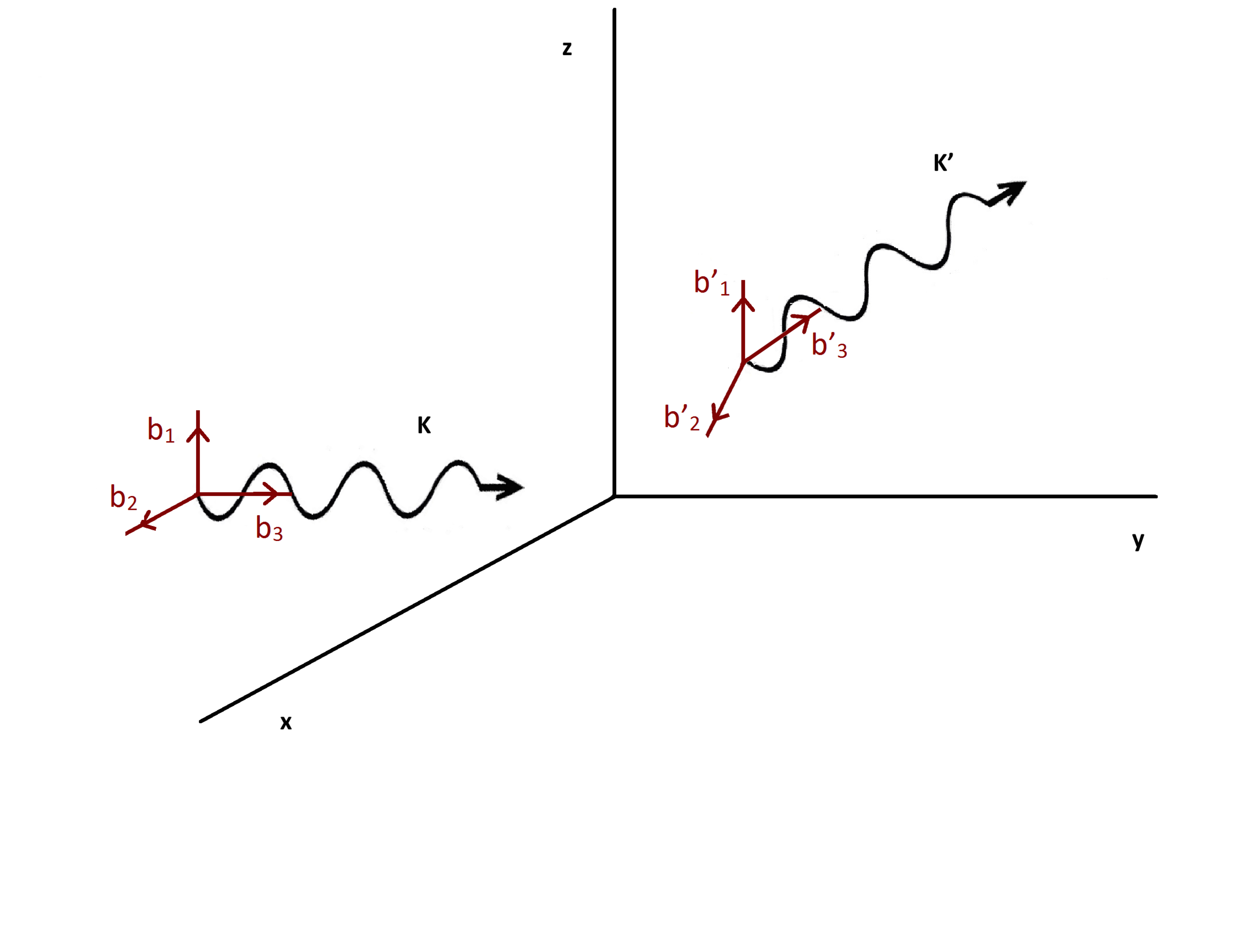} 
\caption{Basis vectors used for determining the Stokes parameters for a Compton scattering process in the 
electron rest frame before (non-primed indices) and after (primed indices) the scattering. 
$\bold{b_3}$ points in the direction of the initial photon wave vector, $\bold{b_2}$ lies in the scattering plane, and $\bold{b_1}$ is normal to that plane. The vector $\bold{b'_3}$ points into the direction of the scattered photon, 
$\bold{b'_1} = \bold{b_1}$, and $\bold{b'_2}$ is perpendicular to the plane of $\bold{b'_1}$ and $\bold{b'_3}$.}
\label{cor}
\end{figure}

We use the Stokes parameters and the non-relativistic Raleigh and relativistic Fano  
scattering matrices to calculate the statistical weight of the scattering and polarization fraction of the scattered photon. 
The Stokes parameters are calculated with the help of two sets of basis vectors (see Figure~\ref{cor}).
The projection of the polarization vector onto the first set of basis vectors allows us to calculate
the polarization angle $\chi_0$, and the Stokes parameters 
$Q_0=\Pi_0 \,I\, cos2\chi_0$, and $U_0=\Pi_0\, I\, sin2\chi_0$
with $\Pi_0$ being the polarization fraction of the incoming photon.

The Stokes parameters before and after scattering (subscripts 0 and 1, respectively) are related via: 
\begin{equation}
\left(
\begin{array}{c}
I_1\\
Q_1 \\
U_1\\
\end{array}\right)
= \mathbf{T_{\rm R/K-N}} 
\left(
\begin{array}{c}
I_0\\
Q_0\\
U_0\\
\end{array}\right).
\end{equation}
The Raleigh scattering matrix is given by \citep{cha60}:
\begin{equation}
\mathbf{T_{\rm R}}\,=\frac{1}{2}r_0^2\left(
\begin{array}{ccc}
1+cos^2\theta & sin^2\theta & 0 \\
sin^2\theta & 1+{\rm cos}^2\theta & 0 \\
0 & 0 & 2cos\theta
\end{array}\right)
 \end{equation}
and the expression for the Fano scattering matrix reads \citep{fan57,mcm61}:
\begin{equation}\label{c}
\mathbf{T_{\rm K-N}}=\frac{1}{2}r_0^2(\frac{\epsilon_1}{\epsilon_0})^2\left(
\begin{array}{ccc}
1+cos^2\theta+\frac{1}{m_ec^2}(\epsilon_0-\epsilon_1)(1-cos\theta) & sin^2\theta & 0 \\
sin^2\theta & 1+{\rm cos}^2\theta & 0 \\
0 & 0 & 2cos\theta
\end{array}\right).
\end{equation}
The Stokes parameters give use the polarization fraction $\Pi_1$ and angle $\chi_1$ after scattering
according to Equations (\ref{s1}) and (\ref{s2}) and we use $\chi_1$ to calculate the polarization 
vector $f'$ of the outgoing photon. We transform the wavevector and polarization vector 
back into the corona frame and the global BL coordinates.
 
For each scattering, we multiply the statistical weight of the photon with a factor $w_1\,=\,I_1/I_0$ encoding the
physics of the scattering process in the rest frame of the scattering electron, and with the kinematical 
factor $w_2\,=\, (1-\beta\cos{(\theta_0)})$ with $\theta_0$ being the angle between the photon's and electron's 
momentum vectors in the coronal rest frame.  The factor $w_2$ reflects the higher likelihood of photon-electron 
head-on collisions compared to photon-electron tail-on collisions. As the factor is frequently omitted 
in simulations of Compton interactions we briefly justify it based on the derivation of the rate of 
Compton scatterings of electrons immersed in an isotropic bath of photons \citep{bic17,blu70,zdz91}. 
In the rest frame of a scattering electron, the number of scatterings is given by:
\begin{equation}
\frac{dN'}{dt'}\,=\,c\,\sigma_{\rm T}\,\int f({\bf p}') d^3 p'
\end{equation}
with $c$ being the speed of light, $\sigma_{\rm T}$ the Thomson cross section, ${\bf p}'$ the momentum of the scattered photons, and $f'({\bf p})'$ the number density of photons per momentum volume element $d^3 p'$. The scattering rate can be transformed into the rest frame of the coronal plasma (undashed variables) noting that $dN$ and $f({\bf p})$ are Lorentz scalars, $dt'\,=\,1/\gamma\,dt$, and $d^3p'\,=\,\gamma(1-\cos{(\theta)})\,d^3p$ with $\beta$ being the electron's velocity in units of the speed of light in the corona rest frame. Assuming isotropic photons in the rest frame of the coronal plasma $f({\bf p})=f(p)$, the scattering rate is:
\begin{equation}
\frac{dN}{dt}\,=\,
c\sigma_{\rm T}\,\int (1-\beta\cos{(\theta)}) f(p) d^3p
\label{rate}
\end{equation}
demonstrating that scatterings with pitch angles $\theta$ contribute with a weight proportional to $1-\beta\cos{(\theta)}$.

\begin{figure}
\epsscale{0.6}
\plotone{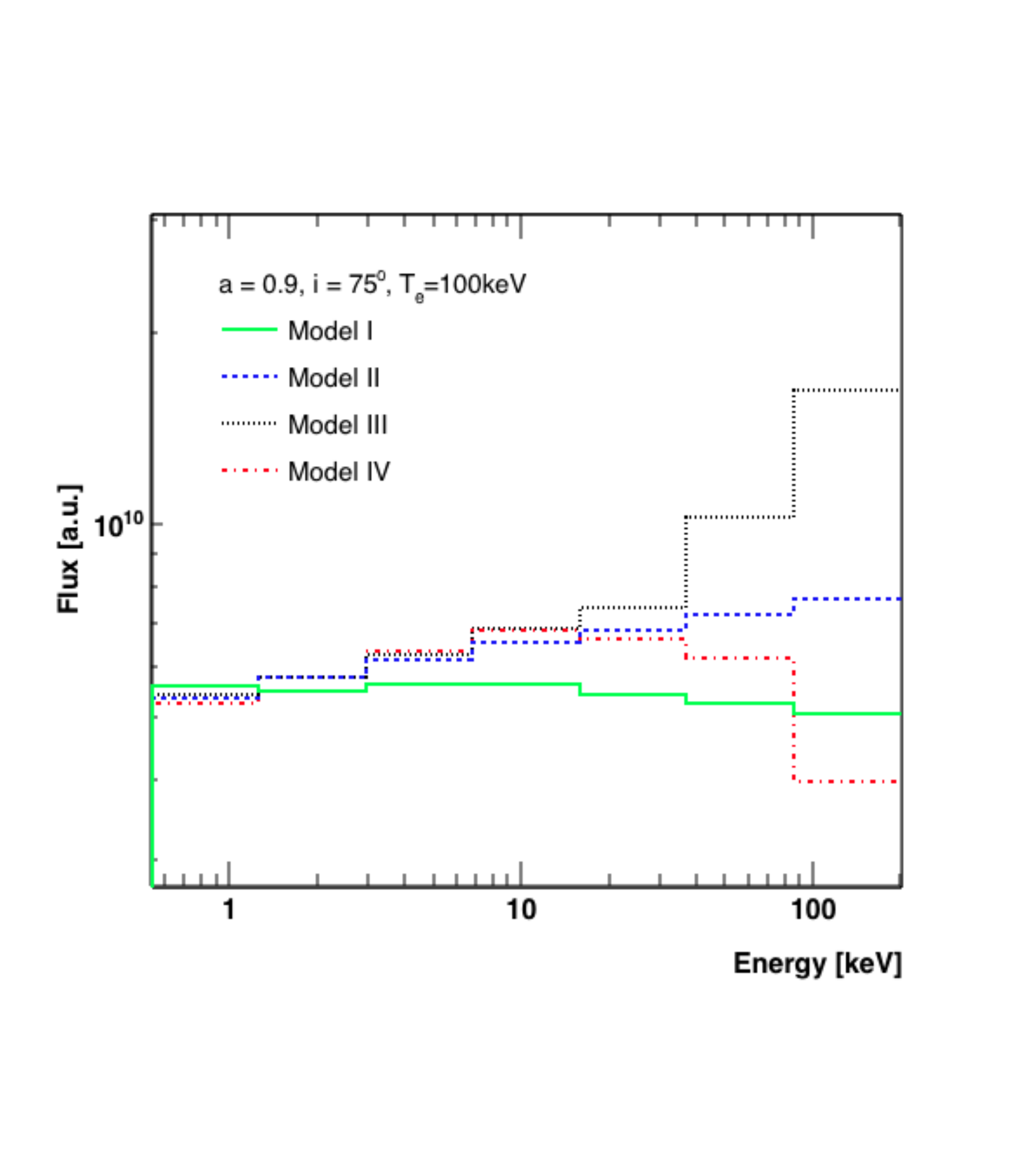} 
\caption{Energy spectrum of the 4 treatments in the coronal scattering. Model (I) is Thomson approximation, (II) is same as (I) but adding the $(1-\beta cos\theta)$ factor, (III) is same as (II) but considering the energy loss of photons, (IV) is same as (III) but using Fano scattering matrix. }
\label{allflux}
\end{figure}

Figure ~\ref{allflux} shows the energy spectra obtained for simulating a spherical corona with $\tau_0=3$ with four different treatments: 
(i) in the Thomson approximation (the photon energy does not change in the rest frame of the scattering electrons),
using the Raleigh scattering matrix and omitting the weighting factor $1-\beta\cos{(\theta)}$;
(ii) same as (i) but including the weighting factor $1-\beta\cos{(\theta)}$;
(iii) as (ii) but accounting for the energy loss of the photon in the rest frame of the scattering electrons,
(iv) as (iii) but using the Fano scattering matrix.  
We see that the weighting factor makes a noticeable difference and changes the 1-10 keV photon index by
$\Delta\Gamma\,\approx0.1$. 
Replacing the Raleigh scattering matrix by the Fano scattering matrix does not noticeably impact the energy spectrum in 1-10 keV; while at higher energies it changes the photon index by $\Delta\Gamma\,\approx0.4$. 

\begin{figure}
\centering
  \begin{tabular}[b]{@{}p{0.45\textwidth}@{}}
    \centering\includegraphics[width=1\linewidth]{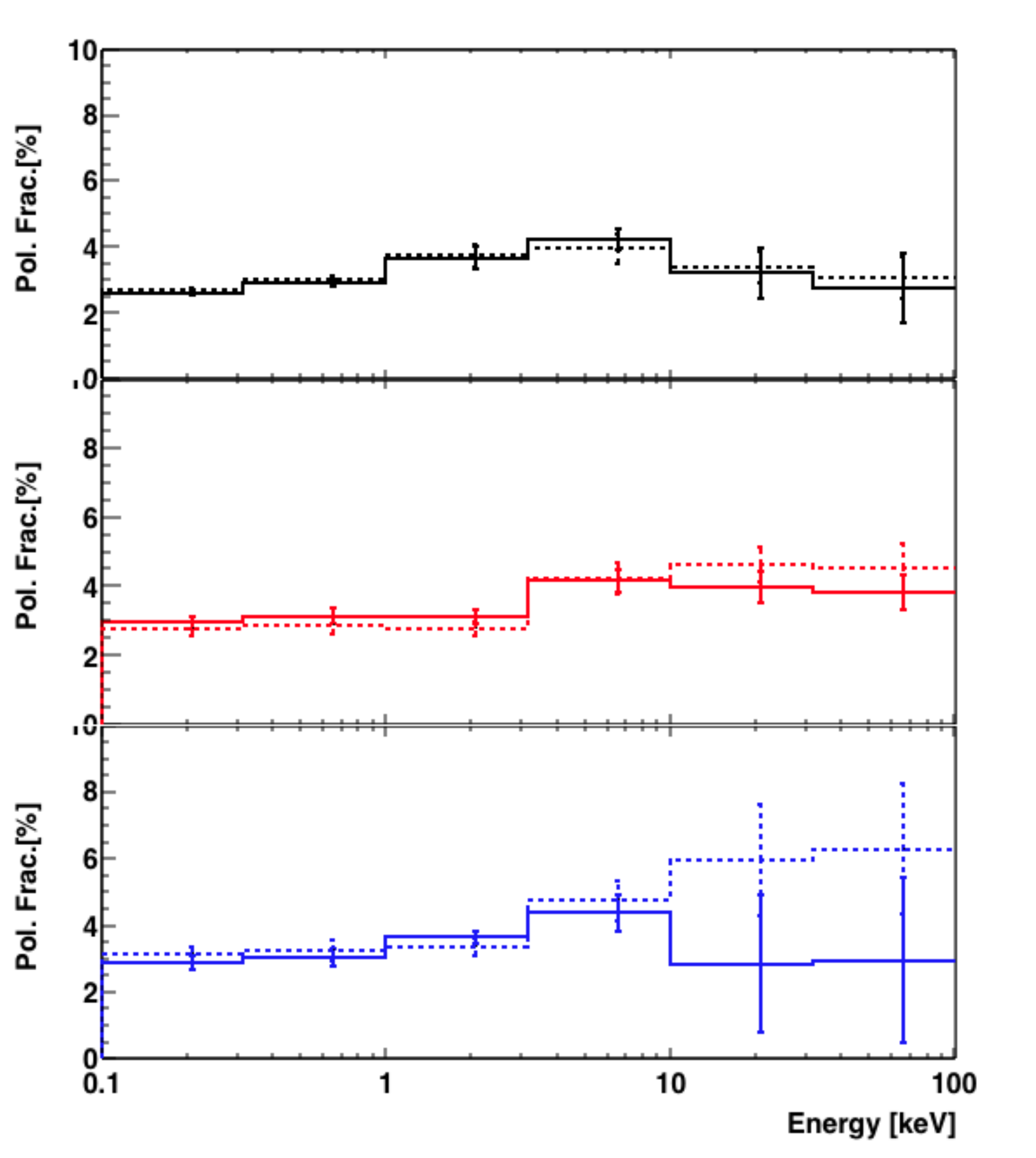} 
    \centering\small (a)
  \end{tabular}%
  \quad
  \begin{tabular}[b]{@{}p{0.45\textwidth}@{}}
    \centering\includegraphics[width=1\linewidth]{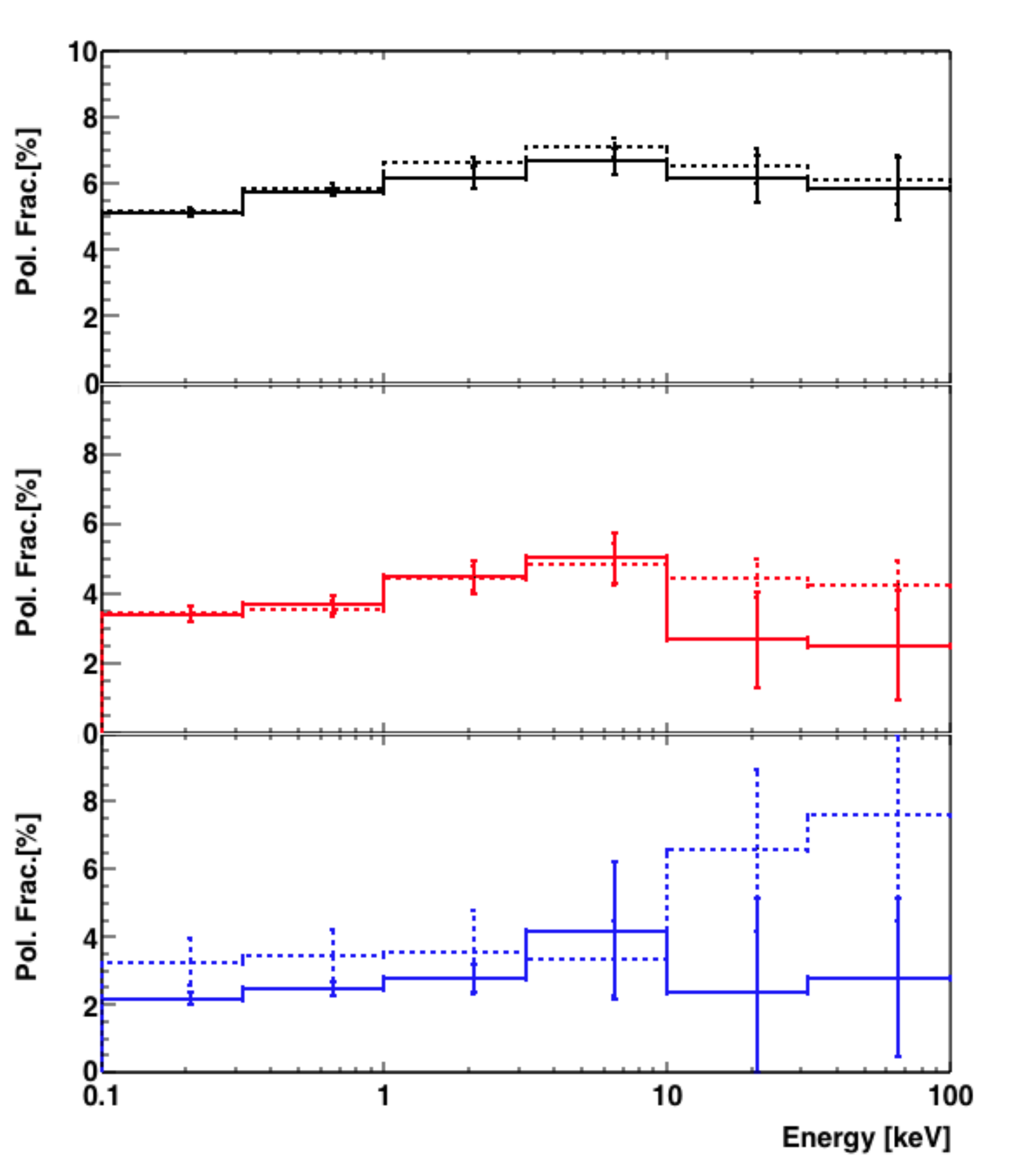} 
    \centering\small (b)
  \end{tabular}
  \caption{Comparison of the polarization fraction energy spectra for the spherical corona (a) and wedge corona (b) 
    calculated with the Thomson cross section, model (iii), (solid line) and the K-N cross section, model (iv), (dashed line) for a black hole  inclination of 75$^{\circ}$. 
    The different panels show the results for different optical depths; left side (spherical corona) from top to bottom: $\tau_0=1.5$/$\sigma=0.12$, $\tau_0=3$/$\sigma=0.24$, $\tau_0=5$/$\sigma=0.39$; right side (wedge corona) from top to bottom: $\tau_0=0.9$/$\sigma=0.16$, $\tau_0=2.3$/$\sigma=0.41$, $\tau_0=4.3$/$\sigma=0.76$.}
  \label{frackn}
\end{figure}

\begin{figure}
\epsscale{1}
 \plotone{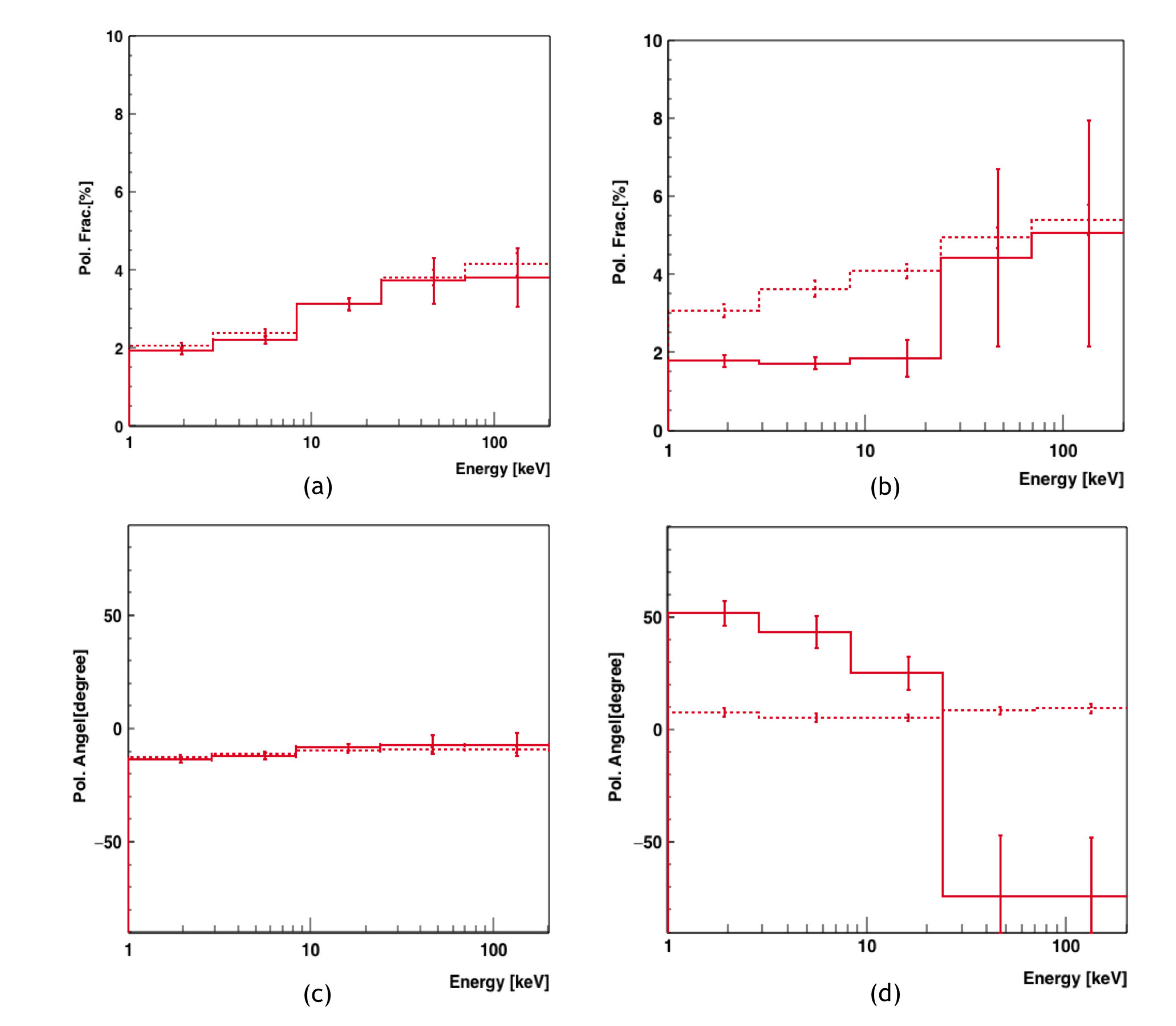} 
\caption{Polarization fraction and angle spectra for the Thomson cross section (solid line) and the KN cross section (dashed line) of the hot accretion disk of a stellar mass black hole seen
at an inclination of $75^{\circ}$. In each panel the optical depths have been chosen to produce the same spectral index in 2-10 keV. (a) Spherical corona with $\tau_0=3,\,\sigma=0.24$ for Thomson and $\tau_0=2.2,\,\sigma=0.17$ for KN with $\Gamma \approx 1$. (b) Wedge corona with $\tau_0=4.,\,\sigma=0.71$ for Thomson and $\tau_0=1.7,\,\sigma=0.3$ for KN with $\Gamma \approx 1.2$.}
\label{HE}
\end{figure}

Figure~\ref{frackn} compares the polarization 
fraction energy spectra for treatments (iii) and (iv) and shows that using the proper 
K-N cross section instead of the Thomson cross section neither impacts the polarization properties 
of the keV photons.
Interestingly, the K-N cross section shows a difference when increasing the energy of the 
seed photons by a factor of 50 (adequate for the accretion disks of accreting stellar mass black holes), see Fig.~\ref{HE}. In these plots corona densities (scattering coefficients) are set to give same spectral index for both treatments (iii) and (iv) in 2-10 keV.
For such high seed photon energies, photons scatter fewer times to get into the X-ray band, 
increasing the importance of each individual scattering. 
The polarization differences are larger for the wedge corona compared to the spherical corona as for the former scatterings are generally 
more important as the corona covers a larger fraction of the inner accretion disk area.
Based on Equ.\ref{a} we expect more pronounced K-N effects at the highest energies. However, the results reveal significant differences at $<$10 keV energies. As the K-N cross section leads to a reduced scattering rate at higher energies, more photons end up in the $<$10 keV band. The higher polarization fraction in K-N scatterings explains the higher polarization of the $<$10 keV photons.

\begin{figure}
\epsscale{1}
\plotone{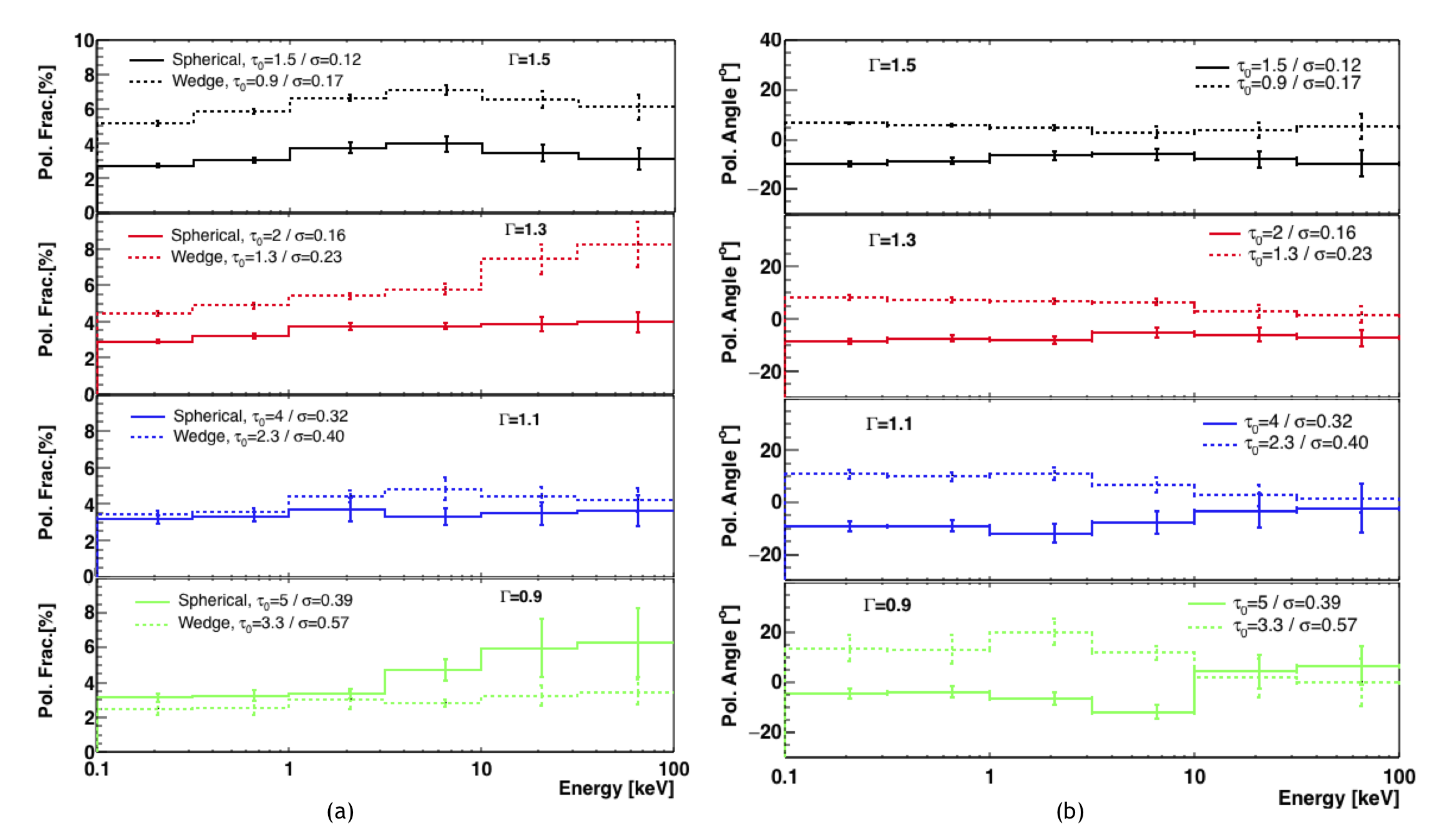} 
\caption{Comparison of the polarization fractions (a) and angles (b) of the spherical corona
    (solid line) and wedge corona (dashed line) for different optical depths, giving for each panel the
    same spectral index (inclination $75^{\circ}$).}
    \label{both}
\end{figure}

Figure~\ref{both} shows the polarization fraction for the same spectral index of the two geometries for different scattering coefficients. We choose corona densities (scattering coefficients) giving the same net 2-10 keV spectral index of the Comptonized emission for the two geometries. This reveals that at $\Gamma>1.1$, lower $\tau_0$/$\sigma$, the wedge corona is more polarized than the spherical corona; while at $\Gamma=1.1$, the two geometries almost have the same polarization fraction; and at steeper spectrum the spherical corona is more polarized at certain  bins. This result shows that the difference between the geometries becomes even somewhat more significant for smaller $\Gamma$-values. The polarization angle of the two models are almost the same at higher energies, while at lower energies for all spectra the two geometry shows opposite polarization direction, Fig.~\ref{both} (b). In the wedge corona photons are influenced by two types of scattering, scattering in the corona and scattering off the disk, while in the spherical corona photons that scatter off the accretion disk are much less than in the wedge corona, so the scattering is dominant by the corona scattering. The angle of polarization for the photons scattered off the disk are in opposite direction of the coronal scattering \citep{sch10}. At lower energies, the two corona shows opposite polarization angle because of the difference in the dominant scattering in the two geometries. While at higher energies mostly coronal scattering is dominated in the two geometries, the polarization angle of the two models are almost the same. In practice, it is hard to distinguish models based on the polarization direction alone, because the orientation of the spin axis of the accretion disk is not well constrained observationally.

\begin{figure}
\epsscale{0.6}
\plotone{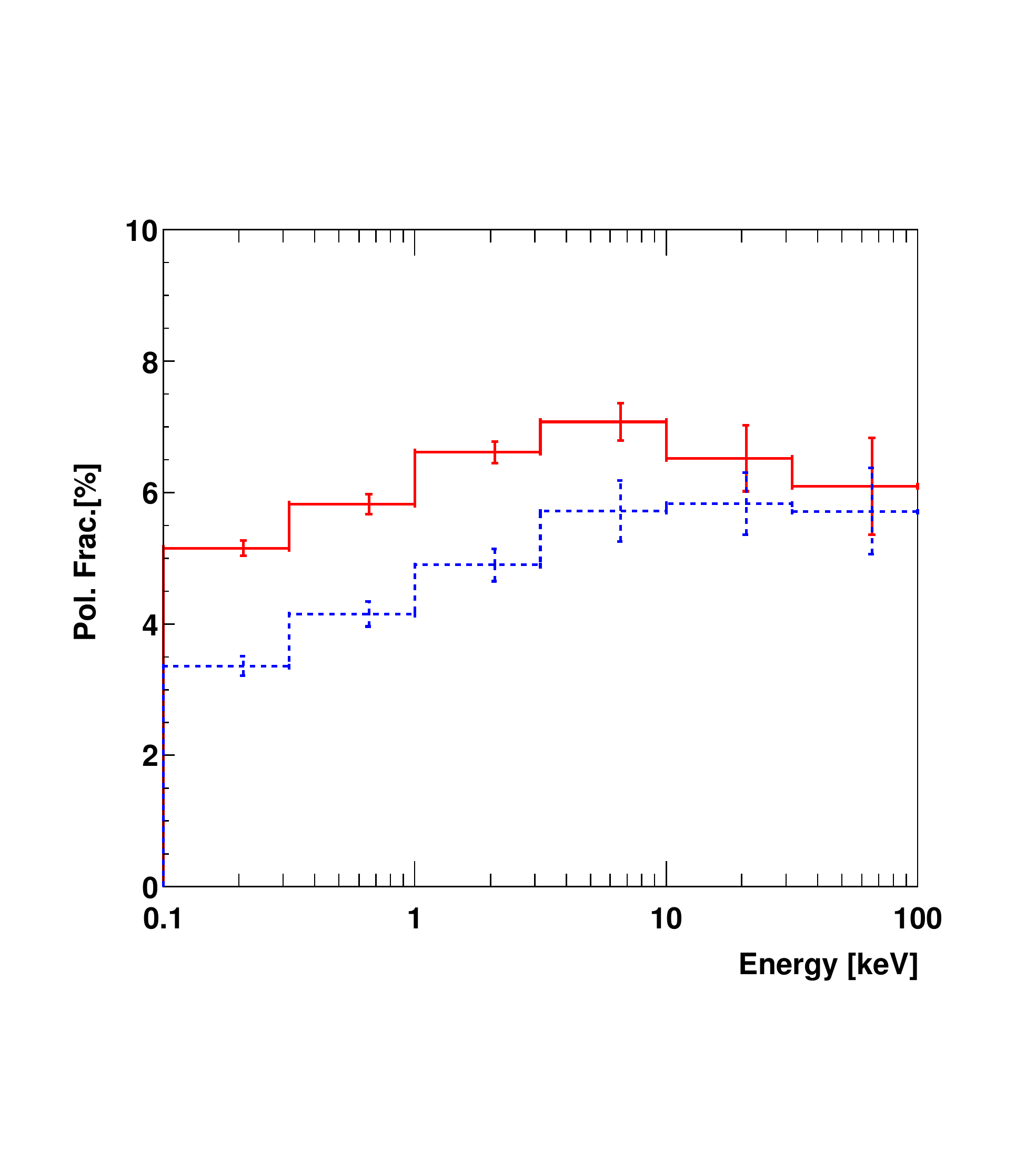} 
\caption{Comparison of the polarization fractions of the wedge corona rotating with ZAMO frame (solid line) and co-rotate with underlying disk (dashed line) for model (iv) at the inclination of $75^{\circ}$.}
    \label{co-zamo}
\end{figure}

As mentioned above, the simulations assume that the wedge corona orbits the black hole with the angular frequency of a ZAMO. As a consequence, the photons originating in the disk will experience Compton scatterings owing to the bulk motion of the coronal plasma relative to the accretion disk. We studied the effect of this bulk motion by running additional simulations with a corona co-rotating with the underlying disk (called {\it Keplerian corona} in the following). Fig.~\ref{co-zamo} compares the polarization fraction of the ZAMO and Keplerian coronas - all other parameters being equal. The results show that ZAMO coronas give higher polarization fractions than Keplerian coronas, owing to the bulk motion of the former. At higher energies, the effect is not noticeable because the large number of scatterings reduce the impact of the first few scatterings on the net polarization signal.

It is important to note that the two coronae cover different portions of the accretion disk with different thermal photon properties. To assess the impact of this point, we performed the same analysis using only seed photons coming from the same portion of the accretion disk for both models. The results was the same as in Fig.~\ref{both}.
In our result, the increase in the polarization fraction when increasing the optical depth/ scattering coefficient is not as clear as \citet{sch10}, Fig. 6 and 15. They increased $\tau$ and adjusted the corona temperature to maintain the same Compton y parameter. Thus, they compare different corona scattering coefficients  for the same energy flux. But in our case, the temperature of the corona is constant and we are comparing polarization for different spectral energies.
\section{Non-thermal plasma}\label{nonth}
In this section assess the impact of a non-thermal power law component of the coronal electron plasma
on the observed polarization properties.  The energy spectra of AGNs and stellar mass black holes in X-ray binaries
often exhibit evidence for the presence of such a component \citep[see][and references therein]{ino14, joh97, cop03, mal09}. 
We assume that the energy spectrum of the thermal electron component is described by the Maxwell-Juttner distribution:
\begin{equation}
\frac{dN}{d\gamma}=\frac{\gamma^2\beta}{\theta K_2(\frac{1}{\theta})} \exp(\frac{-\gamma}{\theta}),
\end{equation}
where $\gamma$ is the Lorentz factor of the electrons, $\theta=k T_{\rm e} / m_{\rm e} c^2$ is the electron temperature
in units of the electron rest mass, and $\beta=v_{\rm e}/c$ is the electron velocity in units of the speed of light. 
The Maxwell-Juttner distribution describes plasmas with temperatures exceeding 100 keV. 
At lower temperatures, the distribution resembles the non-relativistic Maxwell distribution. 
The non-thermal electron component is given by $dN/d\gamma=\xi\gamma^{-p}$ 
with the normalization constant $\xi$  and the power law index $p$. 
(Oriented Scintillation Spectrometer Experiment) observations of NGC 4151, \citet{joh97} 
estimated that $\sim$8\% but less than 15\% of the source power is in the non-thermal electron component.
The results agree with those of \citet{fab17} who studied several {\it NuSTAR} AGN observations and 
estimate that the non-thermal component carries 10\%-30\% of the source power. 
In the following, we assume a non-thermal energy population with $\gamma$-factors between 1 and 1000 
carrying 15\% of the total energy of the coronal electrons.

\begin{figure}
  \centering
 \begin{tabular}[b]{@{}p{0.45\textwidth}@{}}
    \centering\includegraphics[width=1\linewidth]{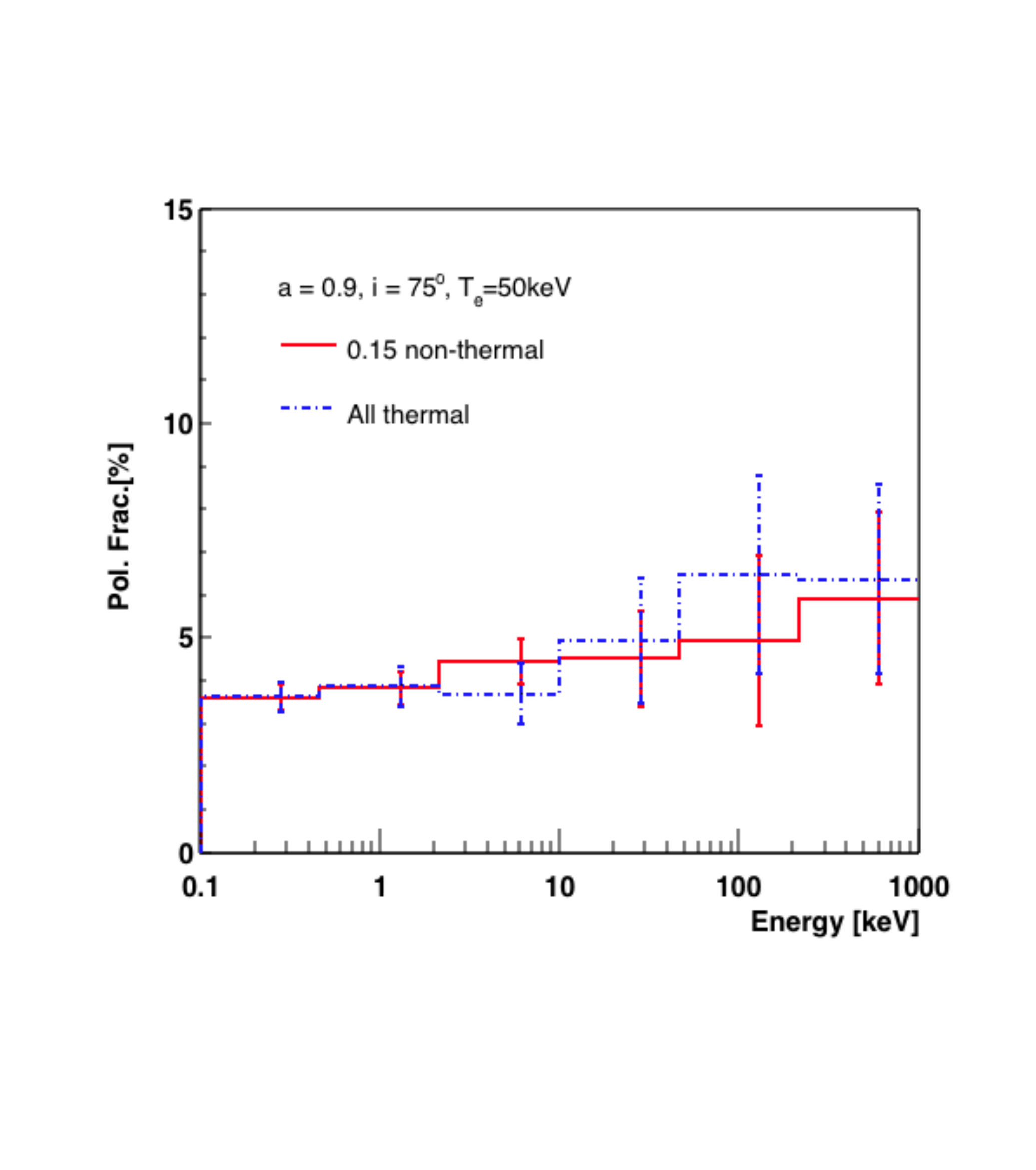} 
    \centering\small (a)
  \end{tabular}%
  \quad
  \begin{tabular}[b]{@{}p{0.45\textwidth}@{}}
    \centering\includegraphics[width=1\linewidth]{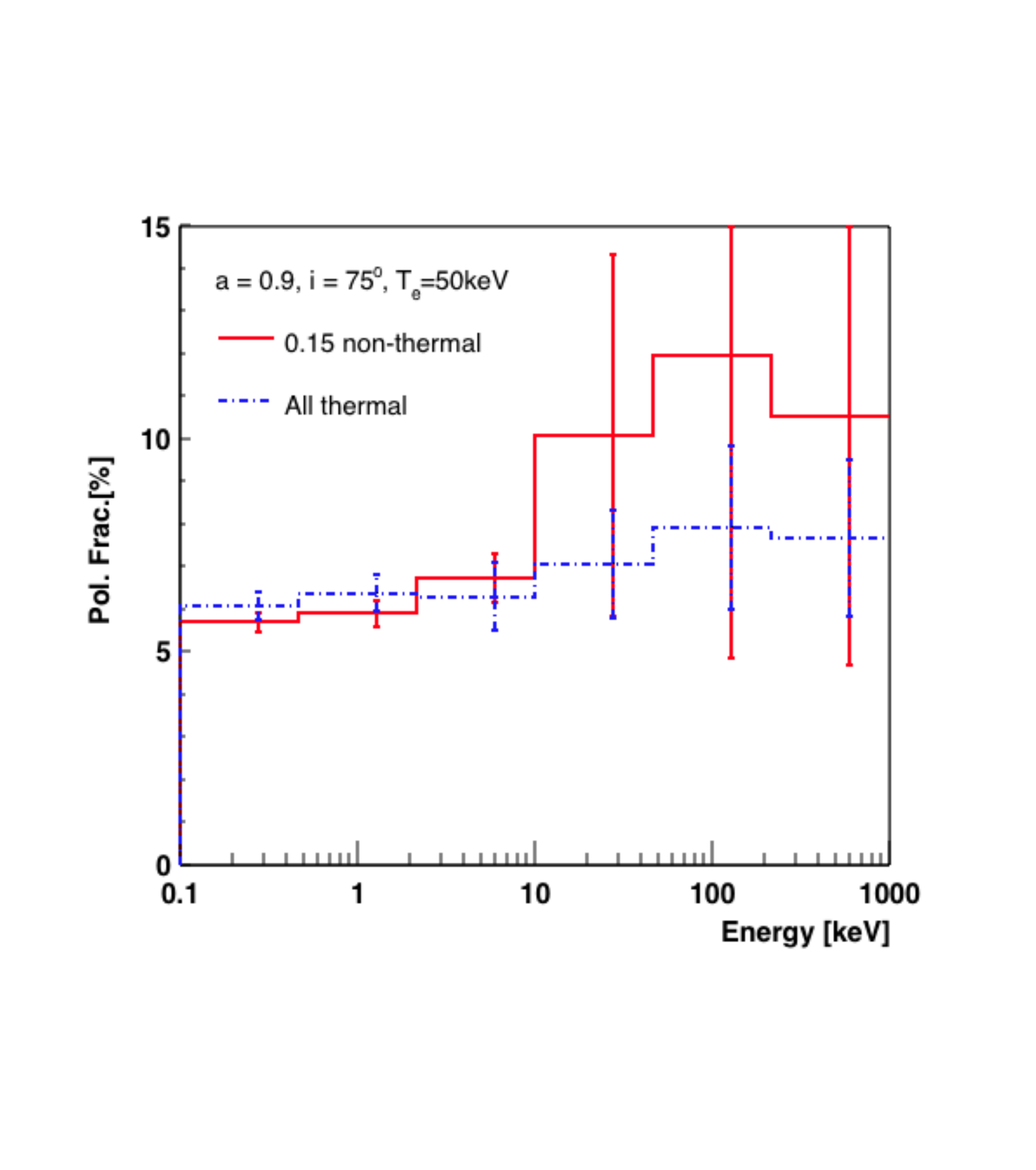} 
    \centering\small (b)
  \end{tabular}
  \caption{Comparison of the polarization predicted for fully thermalized coronal electrons (dot-dashed line) and a thermal plus non-thermal hybrid electron distribution (solid line) for a  spherical corona with $\tau_0=3$ and $\sigma=0.24$ (a) and a wedge corona with $\tau_0=1.9$ and  $\sigma=0.34$ (b).} \label{nonthfrac}
\end{figure}

\begin{figure}
\epsscale{0.6}
\plotone{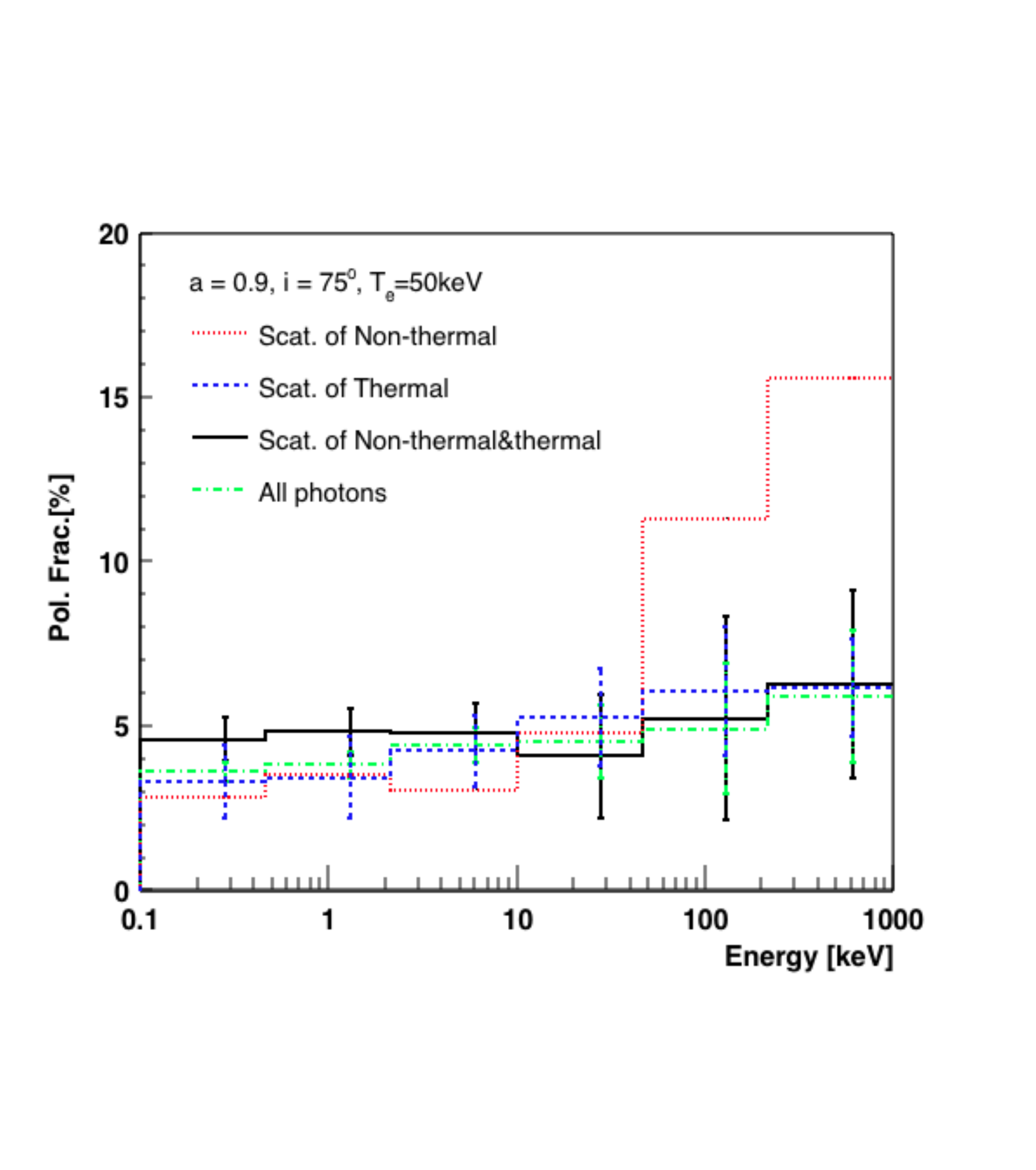} 
\caption{Polarization fractions of different subsets of the photons shown in Fig.~\ref{nonthfrac}(a), i.e.\ 
for photons scattering only off non-thermal electrons (red line), thermal electrons (blue line), the mixture of thermal and non-thermal electrons electrons (black line), and for all photons (green line).
 }
\label{3cont}
\end{figure}

Figure~\ref{nonthfrac} shows the impact of the non-thermal component on the observed polarization signatures.
We assume $p=3$ and a rather low temperature of the thermal component of 50~keV so that we can see 
the impact of the non-thermal electrons at the high-energy end of the observed energy spectra.
For both corona geometries, the addition of a non-thermal electron component carrying 15\% of the energy
barely changes the polarization properties. 
Fig.~\ref{3cont} shows additional details for the spherical corona model. Photons scattering only off non-thermal 
electrons are highly polarized as a small number of scatterings results in a high polarization fraction. 
Photons scattering only off thermal, and off thermal and non-thermal electrons exhibit very similar 
polarization properties. 

\begin{figure}
\epsscale{0.6}
\plotone{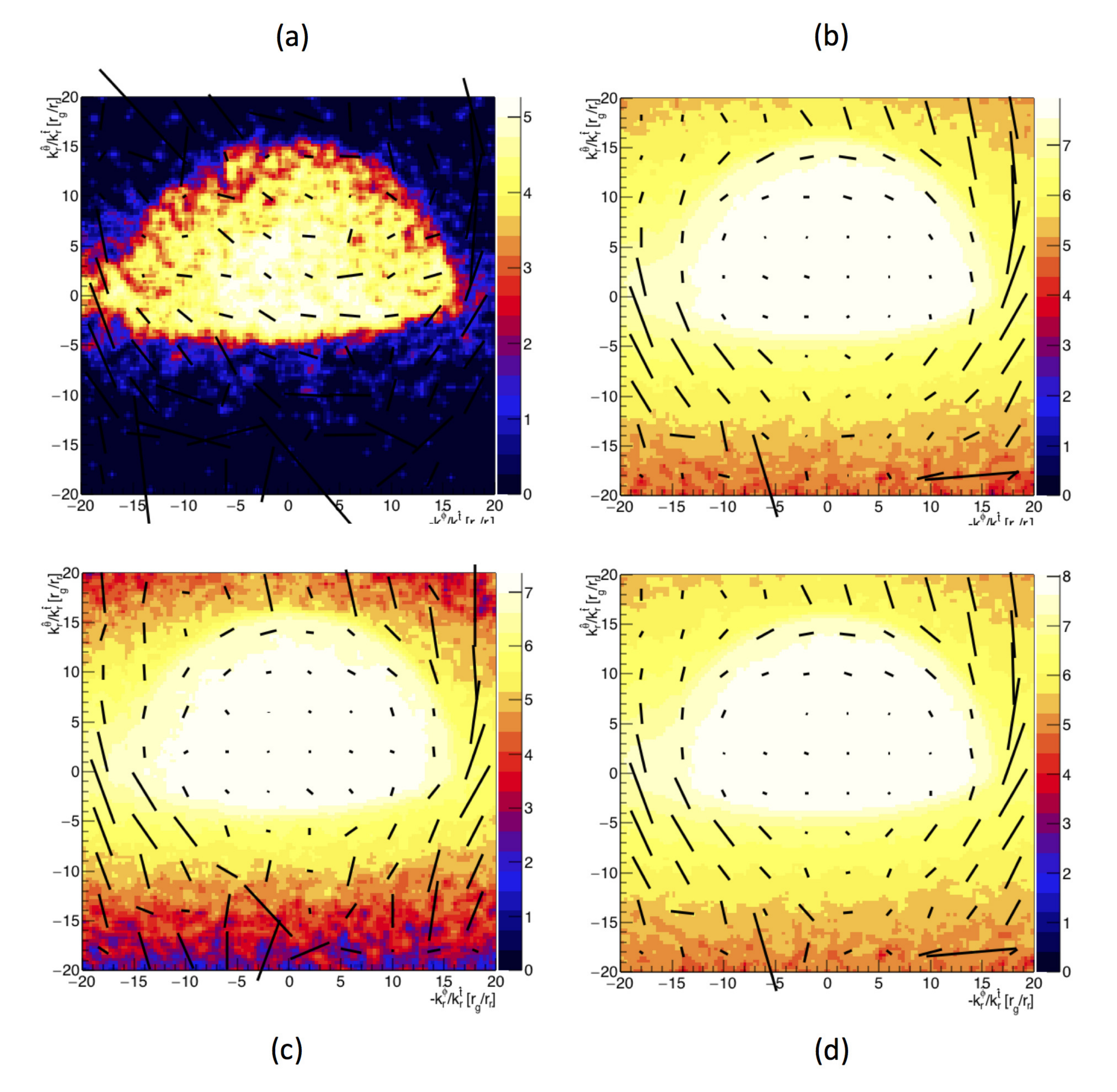} 
\caption{Images of the 1-10 keV photons from Fig.~\ref{3cont} with the color scale showing the intensity in logarithmic scale, 
the length of the black bars is the polarization fraction, and the orientation of the black bars is the direction of the 
preferred electric field of the photons (inclination 75$^{\circ}$). (a) shows photons that scattered only
 off non-thermal electrons, (b) shows photons that scattered only off thermal electrons, (c) shows photons that scattered 
 off two electrons, and (d) shows all photons. }
\label{3contimage}
\end{figure}

Fig.~\ref{3contimage} shows the polarization of photons as function of their arrival direction.
Fig.~\ref{3contimage}(a) shows photons only scattering off non-thermal photons 
with a high polarization fraction (encoded by the length of the black bars) 
when they originate from the inner part of the accretion flow ($<15 r_g$). 
Fig.~\ref{3contimage}(b) shows the much smaller polarization of photons scattering only off thermal electrons, 
and Fig.~\ref{3contimage}(c) shows that the polarization of all photons very much resembles the results of
Fig.~\ref{3contimage}(b). The polarization angle for all images of photons coming from the corona is very comparable and the patterns are very similar. However, the polarization fractions are higher for image (a) explaining the larger bars. Outside of the corona, photons scatter off the disk, so they are more vertically polarized. Inside the corona, the spherical symmetry of the corona and the axial symmetry of the background metric and accretion disk lead to a spherical polarization pattern.

\begin{figure}
  \centering
 \begin{tabular}[b]{@{}p{0.45\textwidth}@{}}
    \centering\includegraphics[width=1\linewidth]{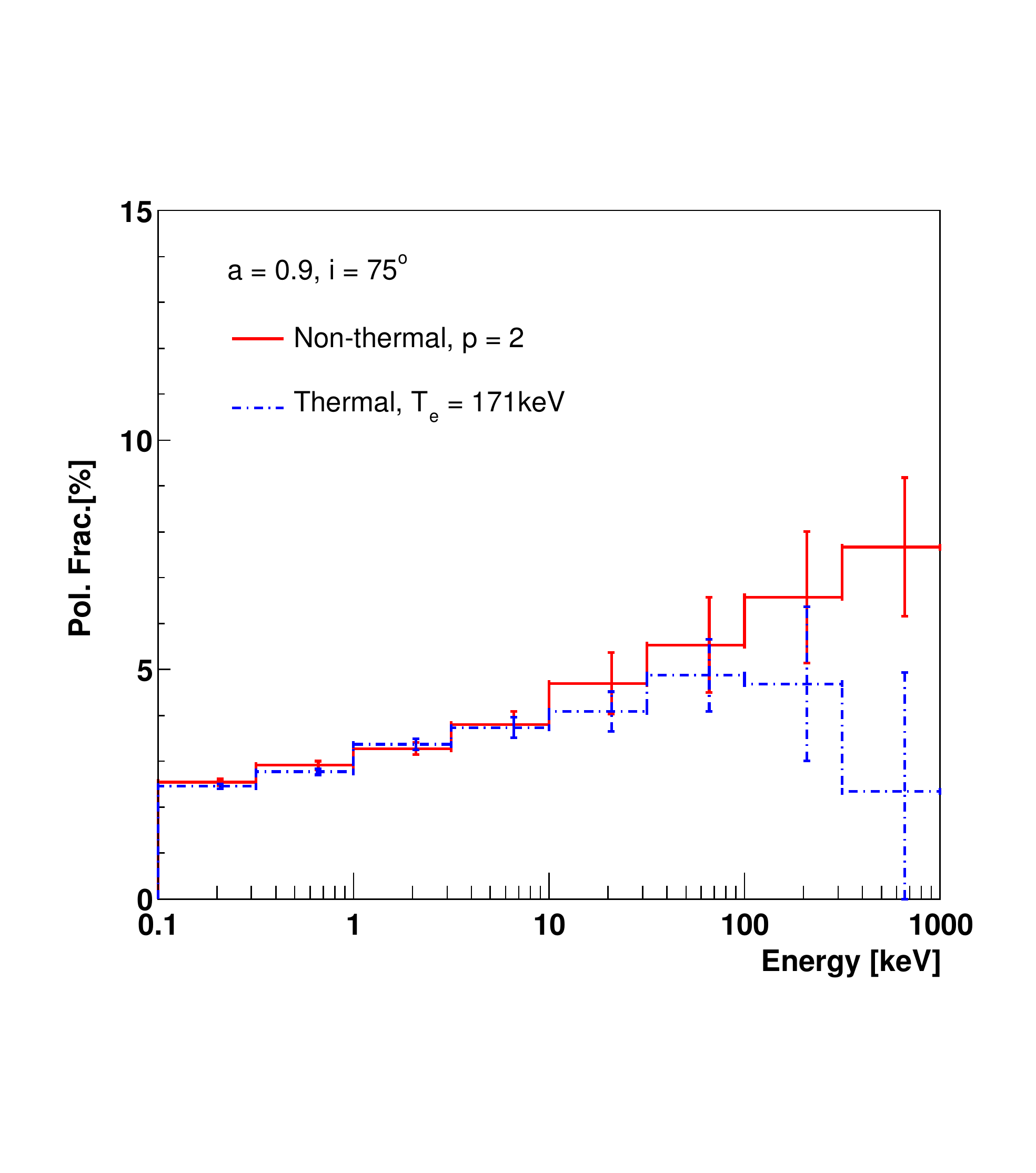} 
    \centering\small (a)
  \end{tabular}%
  \quad
  \begin{tabular}[b]{@{}p{0.45\textwidth}@{}}
    \centering\includegraphics[width=1\linewidth]{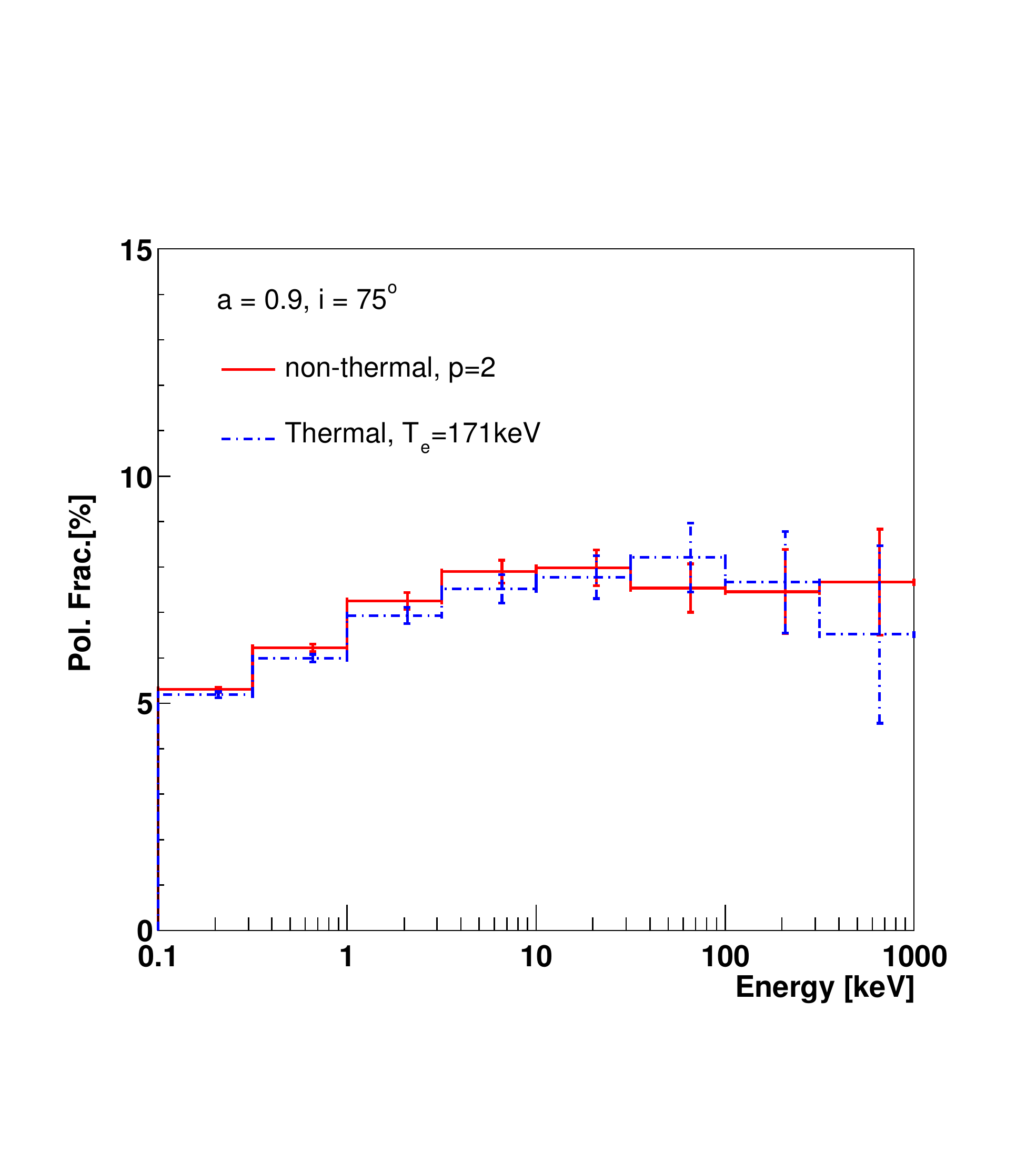} 
    \centering\small (b)
  \end{tabular}
  \caption{Comparison of the polarization of a fully thermal and a fully non-thermal electron energy distribution  for the  spherical (a) and wedge (b) corona geometries.} \label{allnonthpol}
\end{figure}

\citet{ghi93} argue that the coronal plasma may not have sufficient time to thermalize. Figure~\ref{allnonthpol}
compares a thermal model with  $T_{\rm e}= 171$ keV with a non-thermal model with $p=-2$ for $1<\gamma<3$. We choose the same corona temperature and power law index as in \citet{ghi93}, giving the same spectral index in 2-10 keV.
The two models produce almost identical polarization energy spectra with the difference being most pronounced at 
photon energies exceeding 500 keV.
\section{Cyclo-synchrothron seed photons }\label{seed}
Depending on the magnetic field strength, the electrons may loose a good fraction of their energy by emitting 
cyclo-synchrothron photons. In this section, we explore the impact on the X-ray polarization energy spectra. 
We assume an ordered magnetic field of strength $B$ oriented either perpendicular to the disk along the $z$-axis, 
or parallel to the disk in the $x$ and $y$ plane. The cyclo-synchrotron photons are partially circularly and partially
linearly polarized. Since X-ray polarimeters can only measure the linear polarization, we neglect the circular polarization
in the following. As above, we assume the presence of a power law electron component with 
$dN/d\gamma=\xi\gamma^{-p}$ for electrons emitting cyclo-synchrotron photons. 
The synchrotron photons are polarized perpendicular to the magnetic field with a polarization fraction of \citep{ryb79}
\begin{equation}\label{syn1}
\Pi_0=(p+1)/(p+7/3).
\end{equation}
Non-relativistic electron emit cyclotron photons with a polarization fraction of 
\begin{equation}
\Pi_0=\frac{1-cos^2(\theta)}{1+cos^2(\theta)},
\end{equation}
where $\theta$ is the angle between the magnetic field and the line of sight \citep{ryb79}. 
In the following, we consider a scenario in which 50\% of the seed photons are thermal photons from the accretion disk
(polarized according to Chandrasekhar's equation), and 50\% are synchrotron photons from an electron power law distribution with $p=3$ for which Equ.~\ref{syn1} gives a polarization fraction of 75\%. 

Fig~\ref{seed} shows the polarization energy spectra for the spherical corona geometry and the magnetic field 
parallel to the accretion disk. The red line shows the polarization fraction of the synchrotron photons. 
As the synchrotron photons are emitted in the infrared band and only a tiny fraction makes it into the 
X-ray band via a large number of Compton scatterings, the error bars on the polarization fraction and direction
are rather large. Accordingly, the X-ray emission is strongly dominated by the thermal seed photons even
if a substantial fraction of the seed photon energy goes into the synchrotron component.
We obtain the same results for the magnetic field being perpendicular to the accretion disk.
Accounting for the effect of synchrotron self-absorption reduces the overall impact of the synchrotron seed
photons on the observed energy spectra even more.
As cyclotron photons are less energetic and exhibit lower polarization fractions than synchrotron photons, 
we expect that they have a similarly negligible impact on the observed polarization energy spectra. Our result of a negligible impact of the synchrotron seed photons on the emitted energy spectra agrees with the earlier findings of \citep{sch13}.
\section{Simulations of Seyfert \Romannum{1} Galaxy Mrk 335} \label{dis}
In this section we present the results of our code when used to model the {\it NuSTAR} observations of the 
Seyfert \Romannum{1} galaxy Mrk 335 \citet{kee16}. The source harbors a supermassive BH with a mass of 
$M=2.6\times10^7M_{\sun}$ accreting at a rate of $\dot{M}=0.2\dot{M}_{edd}$ \citep{wil15}. 
Fitting the {\it NuSTAR} energy spectrum gives a 2-10 keV photon index of $\Gamma\approx1.9$, and the fit
of the Fe K$\alpha$ line suggests a black hole spin of $a=0.89$ in geometric units and a black hole inclination of $\sim70^{\circ}$. For AGNs, the accretion disk inclination can be inferred from fitting the Fe K-alpha line, or from a combined fit of the flux and polarization energy spectra.

For each corona geometry, we choose three different corona sizes, and adjust the optical depth to recover 
the observed spectral index. Figure~\ref{mrkflux} shows the observed and simulated energy spectra, 
Figure~\ref{mrkpol} shows the polarization fraction, and Table~\ref{table1} lists the model parameters. 
The shaded area in the plots show the energy band that IXPE and XIPE can measure in the near future. In this energy range, there is a clear difference between the two corona geometries, while it is hard to constrain the size. The average polarization fraction of Mrk 335 in the energy of 2-10 keV, The IXPE energy band, as a function of inclination is shown in Fig.~\ref{mrkincli}. In the wedge corona one can clearly see the polarization difference between the  inclinations, while it is hard to compare  in the spherical. The difference between the two geometries is also clear in different inclinations. Mrk 335 is a faint source with the flux of about $10^{-11} erg/cm^2/s$ which will need 2-3 days of IXPE observation for an MDP of 10\%. For brighter sources, e.g. NGC 4151 with one oder of magnitude higher fluxes (e.g. \citet{mar16}), IXPE will achieve an MDP of 10\% in less than two hours and an MDP of 3\% in one day.
 
\section{Discussion}\label{sum}
The results presented above can be summarized as follows: the simulated X-ray polarization energy spectra
depend strongly on the proper treatment of the kinematic effects (change of the photon energy in the rest frame of the
scattering electrons, and relative probability for head on and tail on collisions) and the use of the 
relativistic cross section.
Using the K-N cross section rather than the Thomson cross section increases the 1-10 keV polarization fractions 
by as much as $\sim$3\% for the wedge corona of the hot accretion disk of a stellar mass BH. 
For the colder accretion disks of AGNs, the cross section does not impact the predicted polarization properties
noticeably. The difference between the different corona geometries depends on the optical depth of the coronal plasma,
or, conversely, on the energy spectrum of the observed emission. For high optical depths and hard energy spectra, 
the spherical corona emission is more polarized than the wedge corona emission in the 2-20 keV energy band.
For small optical depths and soft energy spectra, the wedge corona emission exhibits a 1-2\% higher polarization than
the spherical corona over the entire 1-100 keV energy range. The different polarization fractions are accompanied
by differences in the polarization direction.

We find that a non-thermal electron component with about 15\% of the internal electron energy has a negligible
impact on the observed polarization properties. Even a completely non-thermal corona with a small optical depth 
producing the same energy spectrum as a thermal corona shows similar polarization properties as a fully thermalized corona.
Similarly, cyclotron and synchrotron photons do not impact the polarization energy spectra strongly -- 
even when they carry a substantial fraction of the seed photons' luminosity.    
The reason is that these cyclo-synchrotron photons are expected to have too long wavelengths so that only a small fraction
is scattered into the X-ray energy range. Synchrotron photons could have an impact if generated 
with much shorter wavelengths, e.g. as a consequence of magnetic reconnection events in the corona.   

Finally we presented simulations for the Seyfert I galaxy Mrk 335. We predicted the polarization for the two corona geometries and three different corona sizes. Keeping the energy spectrum fixed, the different corona models
predict different polarization fractions and angles. We anticipate that the upcoming IXPE mission will  add valuable
observables for constraining the properties of the inner engine of AGNs - in particular if several complimentary techniques
including spectral, timing, and polarization analyses can all be used for one and the same object.  
\section{Acknowledgments}
BB and HK acknowledge NASA support under grant \#~ NNX14AD19G and NNX16AC42G. JM acknowledges support from  the French Research National Agency (CHAOS project, ANR-12-BS05- 0009, http://www.chaos-project.fr). 
\acknowledgments

\clearpage

\begin{figure}
\epsscale{0.6}
\plotone{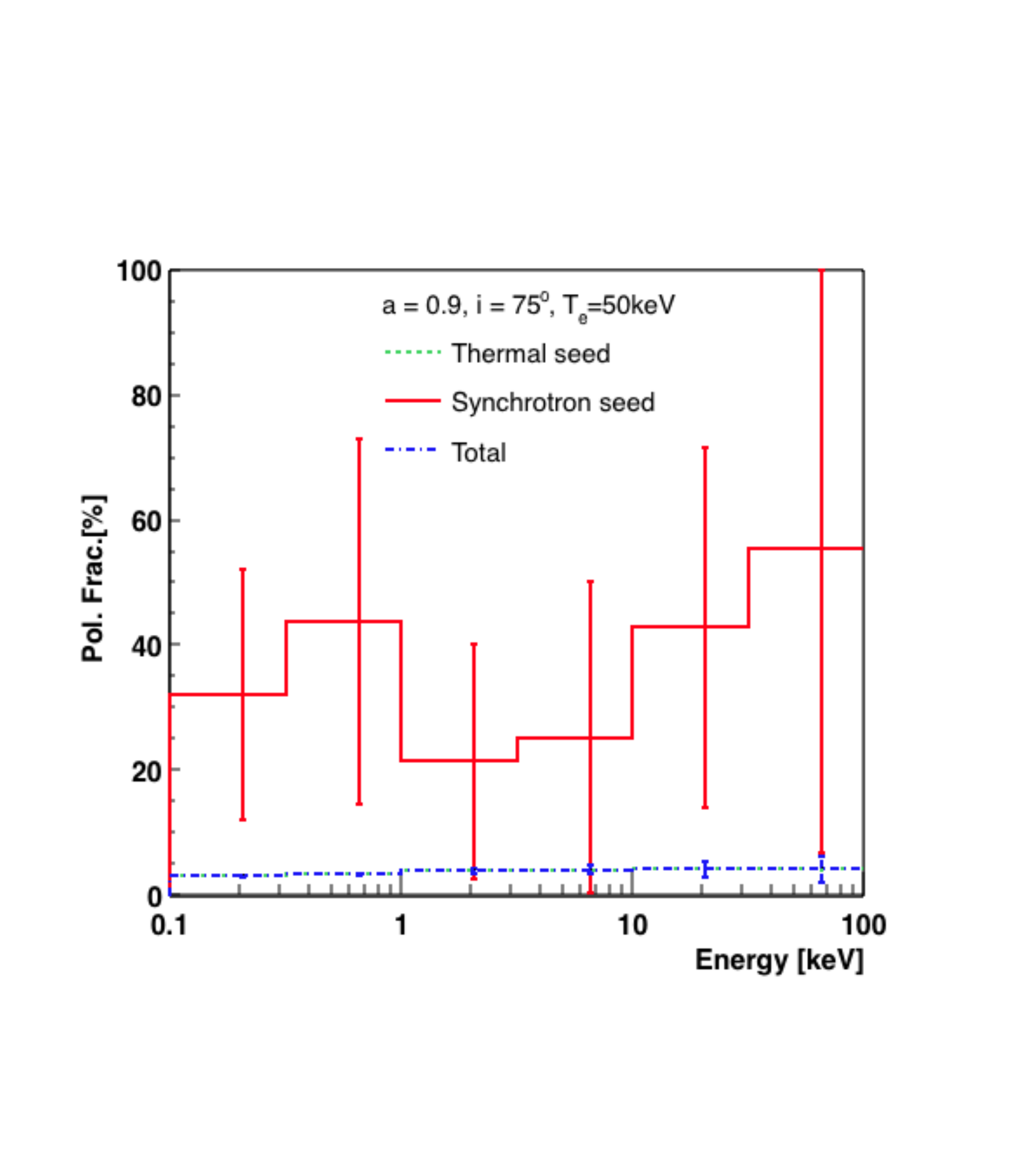} 
\caption{Polarization fractions of the X-ray photons from different seed photons for a spherical corona seen at an inclination of $75^{\circ}$. 
The overall polarization is completely dominated by the photons from thermal seed photons.
The solid line has large error bars because of the very small number of synchrotron photons being scattered into the
0.1-100 keV energy range. }
\label{seed}
\end{figure}

\begin{figure}
\epsscale{0.6}
\plotone{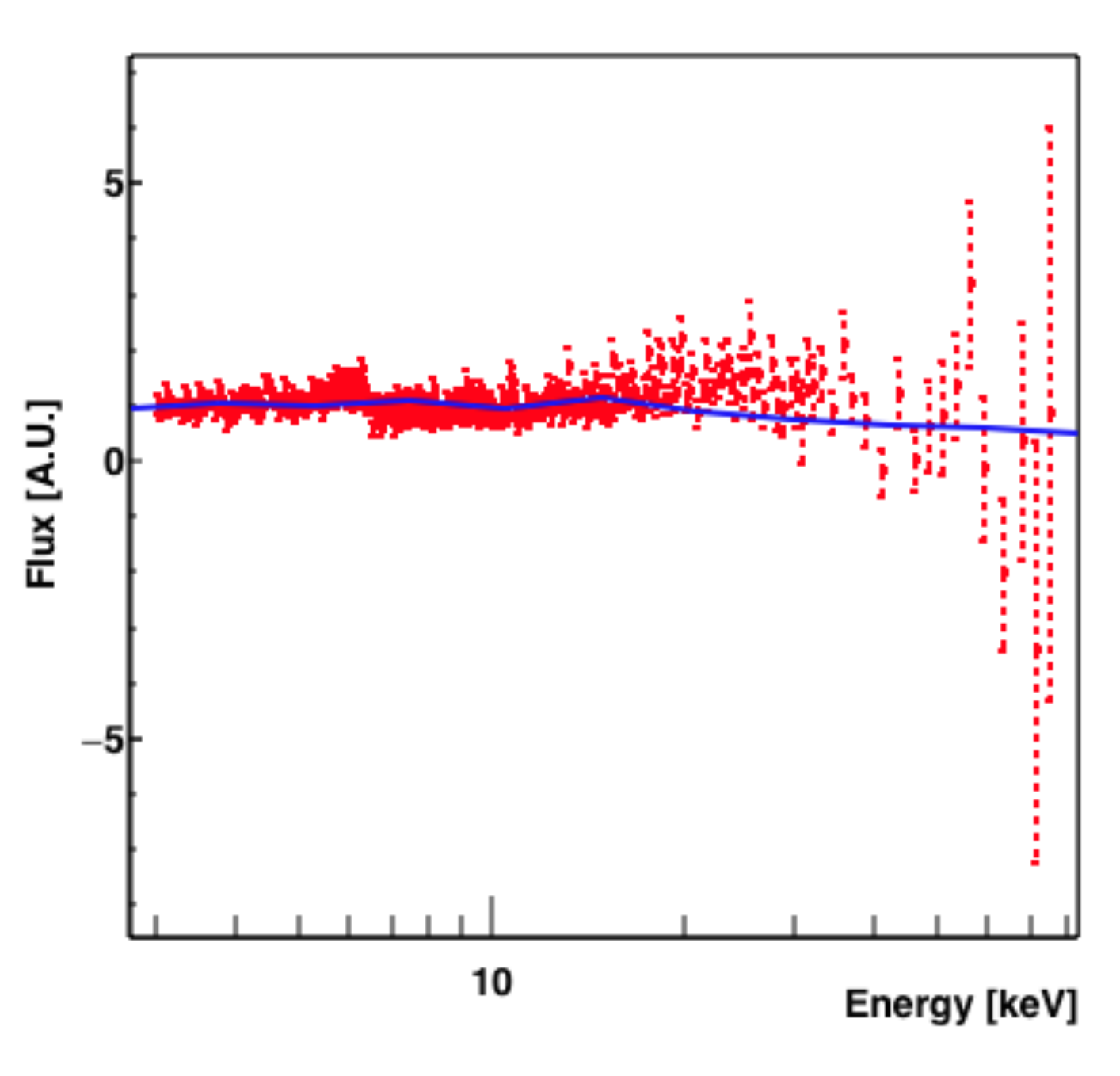} 
\caption{Simulated (solid line) and observed (data points with error bars) spectrum of Mrk 335. The simulations assume
a spherical corona with $R_{\rm edge}\,=\,15\,r_{\rm g}$. }
\label{mrkflux}
\end{figure}

\begin{figure}
  \centering
 \begin{tabular}[b]{@{}p{0.45\textwidth}@{}}
    \centering\includegraphics[width=1\linewidth]{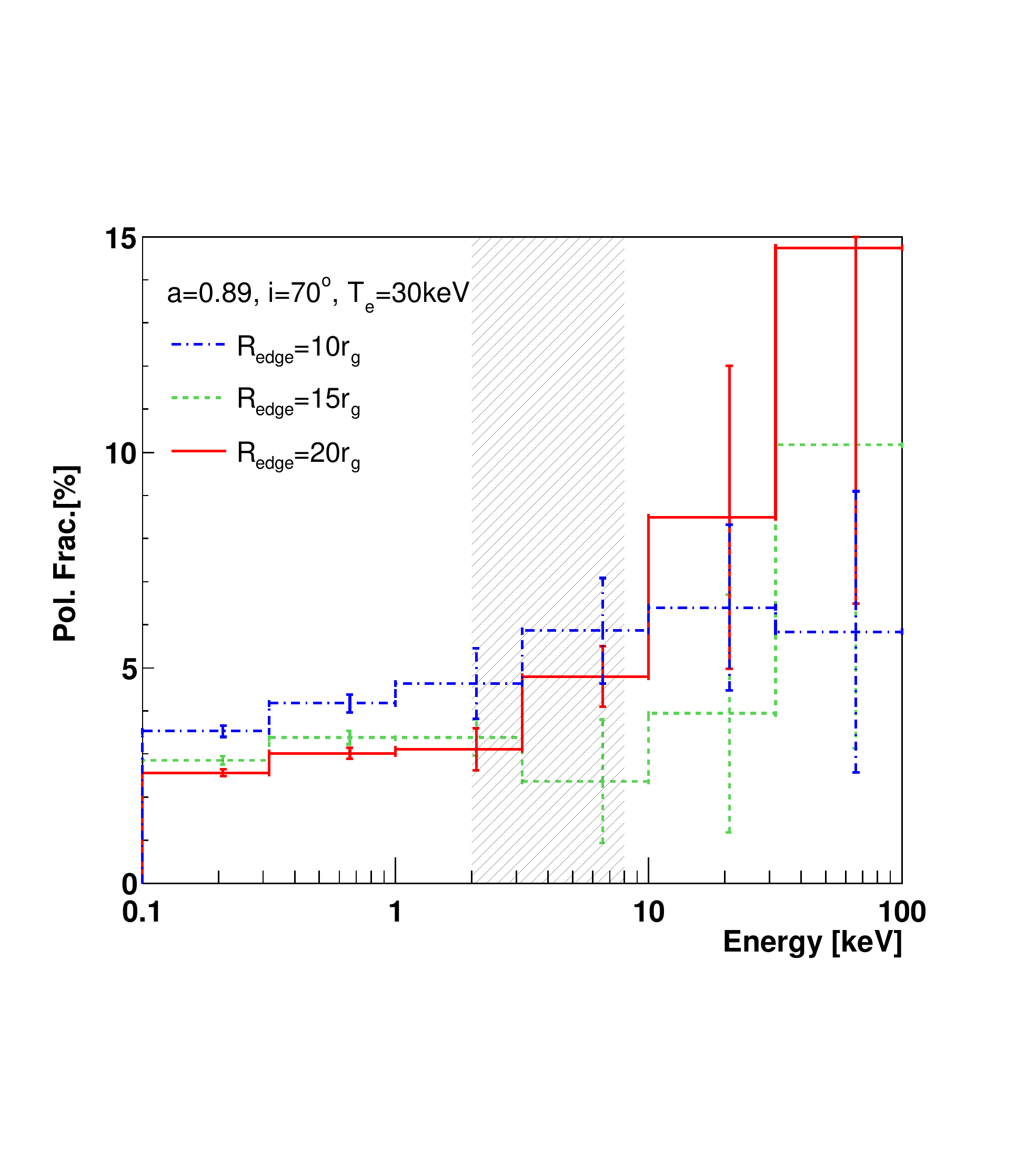} 
    \centering\small (a)
  \end{tabular}%
  \quad
  \begin{tabular}[b]{@{}p{0.45\textwidth}@{}}
    \centering\includegraphics[width=1\linewidth]{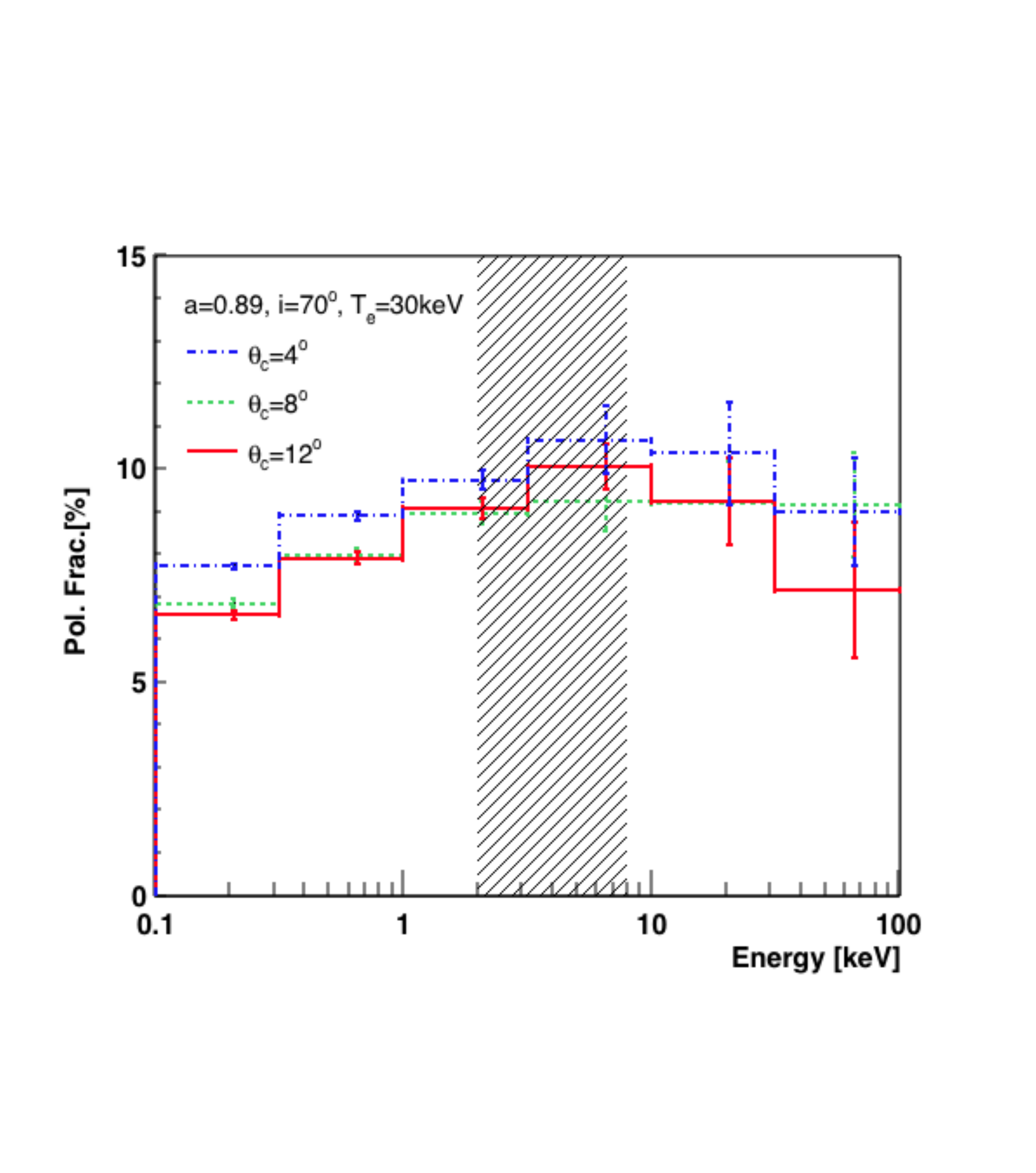} 
    \centering\small (b)
  \end{tabular}
  \caption{Simulated polarization fractions Mrk 335 for coronae of different sizes: (a) Spherical corona, (b) Wedge corona. The shaded area shows the energy range of {\it IXPE}.} \label{mrkpol}
\end{figure}

\begin{figure}
\epsscale{0.7}
\plotone{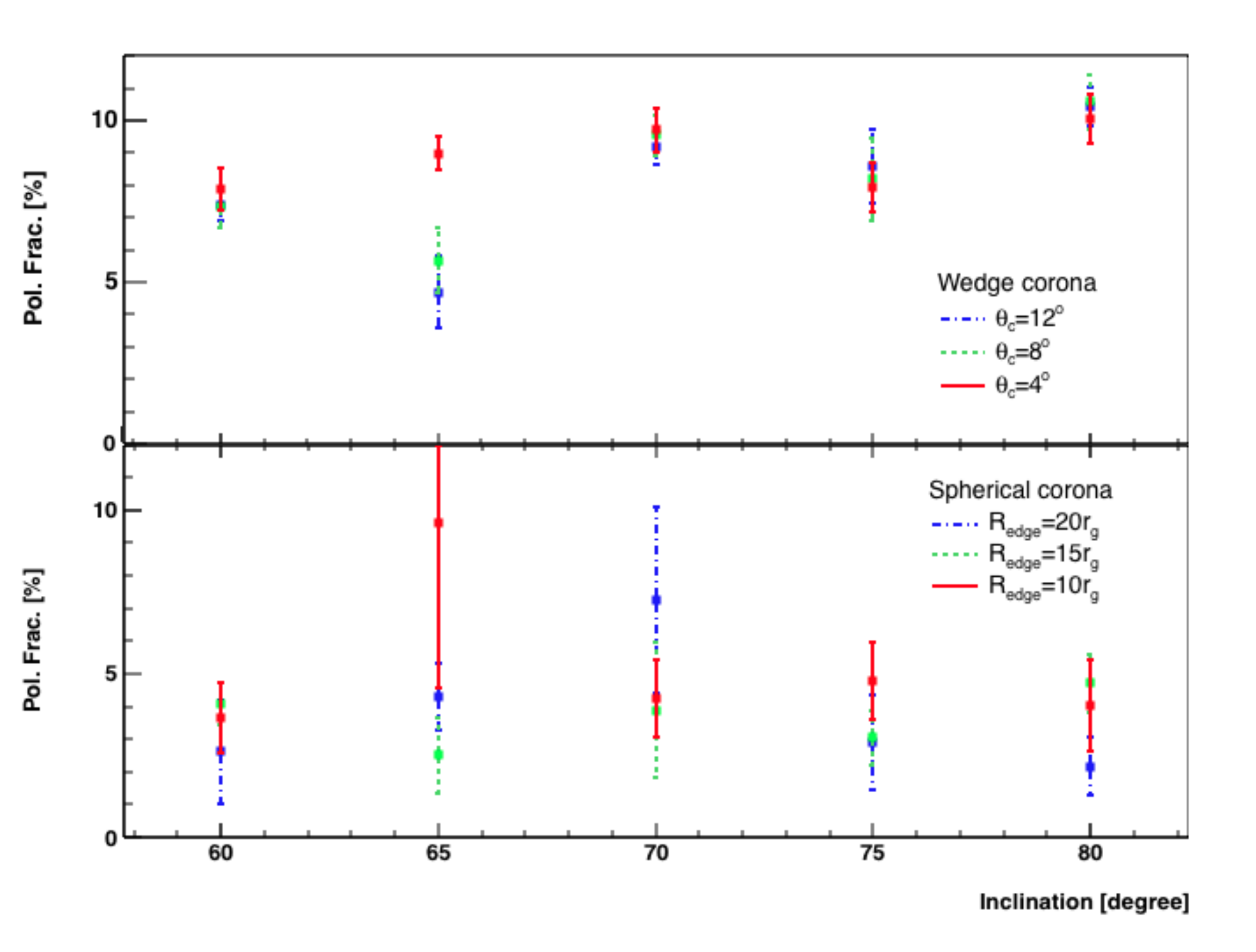} 
\caption{Average Mrk 335 2-10 keV polarization as a function of inclination for the two corona geometries.  }
\label{mrkincli}
\end{figure}

\begin{deluxetable}{ccrrrrrrrrrrcrl}
\tabletypesize{\scriptsize}
\tablecaption{Parameters describing the simulated corona models for Mrk 335\label{table1}}
\tablewidth{0.pt}
\tablehead{
\colhead{Geometry} & 
\colhead{ $R_{edge} \, \left[ r_g \right] $} &
\colhead{$\theta_c \, \left[ ^{\circ} \right]$} & 
\colhead{$\tau_0$} & 
\colhead{$\sigma$} &
\colhead{Temperature (keV)}}
\startdata
 & 10 &NA& 2.3&0.3 & \\ 
Sphere & 15 &NA& 2.5&0.2 & 30 \\
 & 20 &NA& 2.6&0.15 & \\
 \hline
 &NA& 4  & 1.3&0.46 &  \\
Wedge& NA&8 & 1.29&0.23 & 30\\
& NA&12&1.2&0.14 & \\
\enddata
\end{deluxetable}

\end{document}